\documentclass[pra,aps,showpacs,twocolumn,twoside,superscriptaddress]{revtex4}

\usepackage{amsmath,amsfonts,amssymb,caption,color,epsfig,graphics,graphicx,latexsym,mathrsfs,revsymb,theorem,url,verbatim,epstopdf}

\usepackage{hyperref}

\newtheorem{definition}{Definition}
\newtheorem{proposition}[definition]{Proposition}
\newtheorem{lemma}[definition]{Lemma}

\newtheorem{theorem}[definition]{Theorem}
\newtheorem{corollary}[definition]{Corollary}
\newtheorem{conjecture}[definition]{Conjecture}

\newtheorem{remark}[definition]{Remark}
\newtheorem{example}[definition]{Example}
\newtheorem{protocol}[definition]{Protocol}
\newtheorem{question}[definition]{Question}

\def\squareforqed{\hbox{\rlap{$\sqcap$}$\sqcup$}}
\def\qed{\ifmmode\squareforqed\else{\unskip\nobreak\hfil
\penalty50\hskip1em\null\nobreak\hfil\squareforqed
\parfillskip=0pt\finalhyphendemerits=0\endgraf}\fi}
\def\endenv{\ifmmode\;\else{\unskip\nobreak\hfil
\penalty50\hskip1em\null\nobreak\hfil\;
\parfillskip=0pt\finalhyphendemerits=0\endgraf}\fi}
\newenvironment{proof}{\noindent \textbf{{Proof.~} }}{\qed}
\def\Dbar{\leavevmode\lower.6ex\hbox to 0pt
{\hskip-.23ex\accent"16\hss}D}
\makeatletter
\def\url@leostyle{%
  \@ifundefined{selectfont}{\def\UrlFont{\sf}}{\def\UrlFont{\small\ttfamily}}}
\makeatother
\def\bcj{\begin{conjecture}}
\def\ecj{\end{conjecture}}
\def\bcr{\begin{corollary}}
\def\ecr{\end{corollary}}
\def\bd{\begin{definition}}
\def\ed{\end{definition}}
\def\bea{\begin{eqnarray}}
\def\eea{\end{eqnarray}}
\def\bem{\begin{enumerate}}
\def\eem{\end{enumerate}}
\def\bex{\begin{example}}
\def\eex{\end{example}}
\def\bptl{\begin{protocol}}
\def\eptl{\end{protocol}}
\def\bim{\begin{itemize}}
\def\eim{\end{itemize}}
\def\bl{\begin{lemma}}
\def\el{\end{lemma}}
\def\bpf{\begin{proof}}
\def\epf{\end{proof}}
\def\bpp{\begin{proposition}}
\def\epp{\end{proposition}}
\def\bqu{\begin{question}}
\def\equ{\end{question}}
\def\br{\begin{remark}}
\def\er{\end{remark}}
\def\bt{\begin{theorem}}
\def\et{\end{theorem}}

\def\btb{\begin{tabular}}
\def\etb{\end{tabular}}

\newcommand{\nc}{\newcommand}

\def\a{\alpha}
\def\b{\beta}

\def\d{\delta}
\def\e{\epsilon}

\def\t{\theta}

\def\l{\lambda}
\def\m{\mu}
\def\n{\nu}

\def\p{\pi}
\def\r{\rho}
\def\s{\sigma}

\def\ph{\varphi}

\def\ps{\psi}
\def\o{\omega}

\def\L{\Lambda}

\def\Ph{\Phi}
\def\Ps{\Psi}

 \nc{\bA}{{\bf A}} \nc{\bB}{{\bf B}} \nc{\bC}{{\bf C}}
 \nc{\bD}{{\bf D}} \nc{\bE}{{\bf E}} \nc{\bF}{{\bf F}}
 \nc{\bG}{{\bf G}} \nc{\bH}{{\bf H}} \nc{\bI}{{\bf I}}
 \nc{\bJ}{{\bf J}} \nc{\bK}{{\bf K}} \nc{\bL}{{\bf L}}
 \nc{\bM}{{\bf M}} \nc{\bN}{{\bf N}} \nc{\bO}{{\bf O}}
 \nc{\bP}{{\bf P}} \nc{\bQ}{{\bf Q}} \nc{\bR}{{\bf R}}
 \nc{\bS}{{\bf S}} \nc{\bT}{{\bf T}} \nc{\bU}{{\bf U}}
 \nc{\bV}{{\bf V}} \nc{\bW}{{\bf W}} \nc{\bX}{{\bf X}}
 \nc{\bZ}{{\bf Z}}

\nc{\cA}{{\cal A}} \nc{\cB}{{\cal B}} \nc{\cC}{{\cal C}}
\nc{\cD}{{\cal D}} \nc{\cE}{{\cal E}} \nc{\cF}{{\cal F}}
\nc{\cG}{{\cal G}} \nc{\cH}{{\cal H}} \nc{\cI}{{\cal I}}
\nc{\cJ}{{\cal J}} \nc{\cK}{{\cal K}} \nc{\cL}{{\cal L}}
\nc{\cM}{{\cal M}} \nc{\cN}{{\cal N}} \nc{\cO}{{\cal O}}
\nc{\cP}{{\cal P}} \nc{\cQ}{{\cal Q}} \nc{\cR}{{\cal R}}
\nc{\cS}{{\cal S}} \nc{\cT}{{\cal T}} \nc{\cU}{{\cal U}}
\nc{\cV}{{\cal V}} \nc{\cW}{{\cal W}} \nc{\cX}{{\cal X}}
\nc{\cZ}{{\cal Z}}

\nc{\hA}{{\hat{A}}} \nc{\hB}{{\hat{B}}} \nc{\hC}{{\hat{C}}}
\nc{\hD}{{\hat{D}}} \nc{\hE}{{\hat{E}}} \nc{\hF}{{\hat{F}}}
\nc{\hG}{{\hat{G}}} \nc{\hH}{{\hat{H}}} \nc{\hI}{{\hat{I}}}
\nc{\hJ}{{\hat{J}}} \nc{\hK}{{\hat{K}}} \nc{\hL}{{\hat{L}}}
\nc{\hM}{{\hat{M}}} \nc{\hN}{{\hat{N}}} \nc{\hO}{{\hat{O}}}
\nc{\hP}{{\hat{P}}} \nc{\hR}{{\hat{R}}} \nc{\hS}{{\hat{S}}}
\nc{\hT}{{\hat{T}}} \nc{\hU}{{\hat{U}}} \nc{\hV}{{\hat{V}}}
\nc{\hW}{{\hat{W}}} \nc{\hX}{{\hat{X}}} \nc{\hZ}{{\hat{Z}}}

\nc{\hn}{{\hat{n}}}

\def\des{\mathop{\rm des}}
\def\diag{\mathop{\rm diag}}
\def\dim{\mathop{\rm Dim}}

\def\max{\mathop{\rm max}}
\def\min{\mathop{\rm min}}

\def\rank{\mathop{\rm rank}}

\def\sr{\mathop{\rm sr}}
\def\sv{\mathop{\rm sv}}

\def\tr{{\rm Tr}}

\def\sch{\mathop{\rm Sch}}

\def\dg{\dagger}

\def\lf{\lfloor}
\def\rf{\rfloor}
\def\op{\oplus}
\def\ox{\otimes}

\def\ra{\rightarrow}

\newcommand{\bra}[1]{\langle#1|}
\newcommand{\ket}[1]{|#1\rangle}
\newcommand{\proj}[1]{| #1\rangle\!\langle #1 |}
\newcommand{\ketbra}[2]{|#1\rangle\!\langle#2|}
\newcommand{\braket}[2]{\langle#1|#2\rangle}

\newcommand{\norm}[1]{\lVert#1\rVert}
\newcommand{\abs}[1]{|#1|}

\def\Dbar{\leavevmode\lower.6ex\hbox to 0pt
{\hskip-.23ex\accent"16\hss}D}

\begin{document}
\title{Entangling and assisted entangling power of bipartite unitary operations}


\date{\today}

\pacs{03.65.Ud, 03.67.Lx, 03.67.Mn}

\author{Lin Chen}\email{linchen@buaa.edu.cn}
\affiliation{School of Mathematics and Systems Science, Beihang University, Beijing 100191, China}
\affiliation{International Research Institute for Multidisciplinary Science, Beihang University, Beijing 100191, China}
\author{Li Yu}\email{yupapers@sina.com}
\affiliation{National Institute of Informatics, 2-1-2 Hitotsubashi, Chiyoda-ku, Tokyo 101-8430, Japan}

\begin{abstract}
Nonlocal unitary operations can create quantum entanglement between distributed particles, and the quantification of created entanglement is a hard problem. It corresponds to the concepts of entangling and assisted entangling power when the input states are, respectively, product and arbitrary pure states.
We analytically derive them for Schmidt-rank-two bipartite unitary and some complex bipartite permutation unitaries. In particular, the entangling power of permutation unitary of Schmidt rank three can take only one of two values: $\log_2 9 - 16/9$ or $\log_2 3$ ebits. The entangling power, assisted entangling power and disentangling power of $2\times d_B$ permutation unitaries of Schmidt rank four are all $2$ ebits. These quantities are also derived for generalized Clifford operators.
We further show that any bipartite permutation unitary of Schmidt rank greater than two has entangling power greater than $1.223$ ebits. We construct the generalized controlled-NOT (CNOT) gates whose assisted entangling power reaches the maximum. We quantitatively compare the entangling power and assisted entangling power for general bipartite unitaries, and study their connection to the disentangling power. We also propose a probabilistic protocol for implementing bipartite unitaries.
\end{abstract}

\maketitle


\section{Introduction} \label{sec:intro}

In quantum physics, nonlocal unitary operations can create and annihilate entanglement. Bipartite nonlocal unitary operations and entanglement are, respectively, a basic type of operation and a basic type of resource for implementing quantum information processing tasks and studying fundamental problems, such as quantum computing and steering \cite{zmc16}. The bipartite nonlocal unitary operation $U$ on system $A$ and $B$ is a unitary gate that is not the tensor product of any two local unitary gates, i.e., $U\neq V_A \ox W_B$. In other words, $U$ has \textit{Schmidt rank} greater than one.
The entanglement of a bipartite pure state $\ket{\ps}_{AB}$ is defined as the von Neumann entropy $S(\cdot)$ of the reduced density matrix on any one system,
\bea
E(\ket{\psi}_{AB}):=S(\tr_{A} \proj{\psi}).
\eea

In this paper we investigate the following problem: How is the entanglement of a bipartite pure state quantitatively changed under the action of a bipartite nonlocal unitary gate \cite{Nielsen03,lsw09,mkz13,cy15,cy16,cy13,sm10}? Here the state is referred to as the \textit{input state} and contains ancilla systems that are not directly subject to the gate. Bipartite unitaries may create more entanglement than that of the input state. The maximum amount of entanglement increase over all input states is a lower bound of the entanglement cost for implementing bipartite unitaries under local operations and classical communications (LOCC). Our main motivation for studying the entangling capabilities of bipartite unitaries is to try to get insight on the following question, which we think belongs to the class of questions on (ir)reversibility of resources in quantum computation.

\emph{Is there a bipartite unitary such that its entanglement cost is strictly greater than its ability to create entanglement?}

In the following we formalize the notion of the ``ability to create entanglement'' by introducing two types of entangling powers, and their technical definitions are given in Sec. \ref{sec:1a}. [The term ``entanglement cost'' also has a few different definitions; see the text around \eqref{eq:ecbound} and the formalized question stated after it.] The first type of entangling power is when the input state is restricted to a product pure state; and for the second type, the input state is an arbitrary pure state. Both types allow the input state to be on both the systems directly subject to the action of the unitary and some ancillary systems. The two types are respectively called the \emph{entangling power} \cite{Nielsen03} and the \emph{assisted entangling power}. Another quantity we consider is called the disentangling power \cite{lsw09}, which is the maximum amount of entanglement decrease over all input states (allowing ancillary systems) as a result of applying the unitary. These three quantities are some of the most fundamental physical quantities to evaluate the usefulness of bipartite unitaries. Note that we do not discuss another type of entangling power which has also appeared in the literature \cite{zzf00,cgs05}. This is the average output entanglement (under a specific entanglement measure) over Haar random product input states without ancillae.


\begin{table*}
 \begin{tabular}{|c|c|c
 |c|c|}
\hline
$U$ & $\sch U$  & $K_E(U)$ & $K_{Ea}(U)$ & $K_d(U)$
\\\hline
$d_A \times d_B$ unitaries & 2  &
\begin{tabular}{c}
Lemma \ref{le:keu-sr2complex},
\\
Proposition \ref{pp:db=3},
\\
Lemma \ref{le:coarse}.
\end{tabular}
&
\begin{tabular}{c}
    Partial result
    \\
    by Proposition \ref{pp:aep,control}
\end{tabular}
&
\begin{tabular}{c}
$K_{Ea}(U)$
\\
by Lemma \ref{le:u=ut}
\end{tabular}
\\\hline
\begin{tabular}{c}
$d_A \times d_B$
\\
permutation unitaries
\end{tabular}  & 3  &
\begin{tabular}{c}
$\log_2 9 - 16/9$ or $\log_2 3$ ebits
\\
by Proposition \ref{pp:entpower3}
\end{tabular}
&
\begin{tabular}{c}
    Partial result
    \\
    by Proposition \ref{pp:aep,control}
\end{tabular}
     &
\begin{tabular}{c}
$K_{Ea}(U)$
\\
 by Proposition \ref{pp:dx2,sr3}
\end{tabular}
\\\hline
\begin{tabular}{c}
$d_A \times d_B$ complex
\\
permutation unitaries
\end{tabular}  & 3  &
\begin{tabular}{c}
    Partial result
    \\
    by Lemma \ref{le:nqp=220}
\end{tabular}
& \begin{tabular}{c}
    Partial result
    \\
    by Proposition \ref{pp:aep,control}
\end{tabular}
&
\begin{tabular}{c}
$K_{Ea}(U)$
\\
 by Proposition \ref{pp:dx2,sr3}
\end{tabular}
\\\hline
\begin{tabular}{c}
$2 \times d_B$ complex
\\
permutation unitaries
\end{tabular}  & 3  &
\begin{tabular}{c}
   Proposition \ref{pp:cp3}
\end{tabular}
  & \begin{tabular}{c}
    Partial result
    \\
    by Proposition \ref{pp:aep,control}
\end{tabular}
&
\begin{tabular}{c}
$K_{Ea}(U)$
\\
 by Proposition \ref{pp:dx2,sr3}
\end{tabular}
\\\hline
\begin{tabular}{c}
$2 \times d_B$ complex
\\
permutation unitaries
\end{tabular} & 4  &
\begin{tabular}{c}
    $2$ ebits
    \\
    by Proposition \ref{pp:dax2}
\end{tabular} &
\begin{tabular}{c}
    $2$ ebits
    \\
    by Theorem \ref{thm:dax2}
\end{tabular} &
\begin{tabular}{c}
    $2$ ebits
    \\
    by Theorem \ref{thm:dax2}
\end{tabular}
\\\hline
GCNOT & 2  &
\begin{tabular}{c}
    $1$ ebit
    \\
    by Proposition \ref{pp:gcnot}
\end{tabular}
&
\begin{tabular}{c}
    $1$ ebit
    \\
    by Proposition \ref{pp:gcnot}
\end{tabular}  &
\begin{tabular}{c}
    $1$ ebit
    \\
    by Proposition \ref{pp:gcnot}
\end{tabular}
\\\hline
generalized Clifford operators & no requirement  & Proposition \ref{prop:Clifford} & Proposition \ref{prop:Clifford} &  Proposition \ref{prop:Clifford}
\\\hline
\begin{tabular}{c}
\begin{tabular}{c}
  $d_A\times d_B$
  \\
  permutation unitaries
\end{tabular}

\end{tabular}
& greater than two
&
\begin{tabular}{c}
$>1.223$ ebits
\\
by Proposition \ref{pp:>1ebit}
\end{tabular}
&
\begin{tabular}{c}
$>1.223$ ebits
\\
by \eqref{eq:ecbound} and
\\
Proposition \ref{pp:>1ebit}
\end{tabular}
&
\begin{tabular}{c}
$>1.223$ ebits
\\
by \eqref{eq:kdu=keau} and
\\
Proposition \ref{pp:>1ebit}
\end{tabular}
\\\hline
\begin{tabular}{c}
  a family of
    \\
    non-controlled unitaries
\end{tabular}
& no requirement  & Proposition \ref{pp:gs} & ?
&
\begin{tabular}{c}
  $K_{Ea}(U^\dg)$
    \\
by \eqref{eq:kdu=keau}
\end{tabular}
\\\hline
 \end{tabular}
    \caption{\label{tab:result}
List of the main results of this paper in terms of the type of bipartite unitary $U$. The symbols $\sch U$, $K_E(U)$, $K_{Ea}(U)$, and $K_d(U)$ represent, respectively, the Schmidt rank of $U$, the entangling power of $U$,  the assisted entangling power of $U$, and the disentangling power of $U$. The generalized CNOT (GCNOT) gate is defined in Sec. \ref{subsec:gcnot}. The ``?'' means unknown.
 }
 \end{table*}

To investigate our problem, we study the above three quantities in terms of 
some classes of bipartite unitaries. They include the Schmidt-rank-two bipartite unitaries, the bipartite complex permutation unitaries of Schmidt rank three or four, the generalized CNOT gates, and the bipartite generalized Clifford operators. The importance of these gates is summarized as follows. Bipartite unitaries of Schmidt rank two or three are locally equivalent to
controlled unitary operators \cite{cy13,cy14,cy14ap}. They include the basic ingredients of quantum computing such as CNOT gates and controlled-phase gates. The controlled unitary can be implemented with LOCC and a maximally entangled state \cite{ygc10}, and is the mostly realizable class of nonlocal unitaries by experiments. The equivalence between bipartite and controlled unitaries has also been used to evaluate the delocalization power of bipartite unitaries \cite{sm10}.
As the investigation of bipartite unitaries of greater Schmidt rank is more involved, we focus on the permutation unitary gates. They have a simpler structure than that of arbitrary unitaries and contain experimentally realizable gates such as the SWAP gate. Any bipartite permutation unitary of Schmidt rank three can be implemented using LOCC and two ebits \cite{cy16}. On the other hand, a protocol for implementing bipartite permutation unitaries of any Schmidt rank $r$ has been given, by using $O(r)$ ebits of entanglement and $O(r)$ bits of classical communication \cite{cy16}. The Clifford gates are central for the field of quantum error correction \cite{Gottesman97}, and are interesting for many other topics in quantum information theory.

Our main results are concluded in Table \ref{tab:result} and introduced as follows.
We analytically derive the entangling power of Schmidt-rank-two unitaries, and the results are mainly presented in Lemma \ref{le:keu-sr2complex} and Proposition \ref{pp:db=3}.
In Proposition \ref{pp:entpower3}, we show that the entangling power of bipartite permutation unitary gates of Schmidt rank three can only take one of two values: $\log_2 9 - 16/9$ or $\log_2 3$ ebits. The result is counter-intuitive because one may expect that the entangling power depends on the gate more strongly. We are not aware of a similarly large family of bipartite unitary gates that have exactly two distinct values of entangling power. We analytically construct the gates for the value $\log_2 9 - 16/9$. The value $\log_2 3$ is the upper bound of entangling power of all Schmidt-rank-three bipartite unitaries. Next, we show in Proposition \ref{pp:dax2} that the entangling power of any $2\times d_B$ complex bipartite permutation unitary of Schmidt rank four is 2 ebits, and in Proposition \ref{pp:>1ebit} that any bipartite permutation unitary of Schmidt rank greater than two has entangling power greater than $1.223$ ebits. So permutation unitaries generally have a stronger entangling power and assisted entangling power than that of arbitrary bipartite unitaries, since the latter could approach zero.
Third, we construct the notion of a generalized CNOT (GCNOT) gate and study its entangling power in Proposition \ref{pp:gcnot}. The GCNOT gate has the maximum entangling and assisted entangling power among Schmidt-rank-two bipartite unitaries of high dimensions. So the GCNOT gate plays the same role as the CNOT gate does in the two-qubit unitary gates. Fourth, we construct the notion of generalized Clifford operators and derive their entangling power, assisted entangling power and disentangling power in Proposition \ref{prop:Clifford}. It turns out that they are all equal to the Schmidt strength defined in \cite{Nielsen03} and \eqref{eq:schmidt_strength}.

Other results in Table \ref{tab:result} are introduced in sections. Below we introduce the discussed quantities in terms of their physical meaning and mathematical formulation.

\subsection{Definitions and physical meanings}\label{sec:1a}

The entangling power of a bipartite unitary $U$ acting on the Hilbert space $\cH$ of systems $A,B$ is defined as \cite{Nielsen03}
\bea\label{eq:K_e}
K_E(U) := \max_{\ket{\alpha},\ket{\beta}} E(U(\ket{\alpha}\ket{\beta})).
\eea
Here $\ket{\alpha}$ and $\ket{\beta}$ are pure states on system $A R_A$ and $B R_B$, respectively, $R_A$ and $R_B$ are local ancillae, and the $E$ is the von Neumann entropy of the reduced density matrix on one of the two systems $A R_A$ and $B R_B$. So $\ket{\a,\b}$ and $U\ket{\a,\b}$ are bipartite states. For two bipartite unitaries $U,V$, both acting on $\cH$, we have
$
K_E(U\ox V) \ge K_E(U) + K_E(V).
$
From \cite{Nielsen03}, the collective use of $U,V$ might have a stronger entangling power than the sum of that of $U$ and $V$. This can even happen when $U=V$. Thus, the $K_E$ is not, in general, weakly or strongly additive  \cite{Nielsen03}. This is analogous to the superadditivity of various types of capacities of quantum channels \cite{sy08,lwz09}.

The entangling power needs a product state as the input state, so we do not need entanglement as the initial resource. This is a more efficient way from the point of view of experiments, because entanglement is usually hard to realize in a laboratory. On the other hand from the theoretical point of view, adding the entanglement as an initial resource may increase the entanglement that can be generated by the bipartite unitary. For this purpose we introduce the assisted entangling power.  It also gives a lower bound for the entanglement cost under LOCC. The \emph{assisted entangling power} of a bipartite unitary $U$ is defined as
\bea\label{eq:K_ea}
K_{Ea}(U) := \sup_{\ket{\psi}}
\bigg(
E(U(\ket{\psi}))-E(\ket{\psi})
\bigg).
\eea
Here $\ket{\psi}$ is a bipartite pure state on the systems $A R_A$ and $B R_B$, $R_A$ and $R_B$ are local ancillae, and the $E$ is the von Neumann entropy of the reduced density matrix on one of the two systems $A R_A$ and $B R_B$.
The assisted entangling power has been discussed in the name of ``entangling capacity'' \cite{lhl03}, and another definition without ancillae is also discussed in \cite{lhl03}. On the other hand, the quantity $K_{\Delta E}(U):= \sup_{\ket{\psi}} \vert E(U(\ket{\psi}))-E(\ket{\psi})\vert$ defined in \cite{Nielsen03} is lower bounded by $K_{Ea}(U)$. From the definition of weak additivity in \cite{Nielsen03}, and the definitions of $K_E$ and $K_{Ea}$, it can be deduced that if $K_{Ea}=K_E$ for some class of bipartite unitaries, then $K_E$ is weakly additive for them. It is shown in \cite{Nielsen03} that $K_E$ is strictly subadditive for some two-qubit unitaries, thus $K_E(U)<K_{Ea}(U)$ for some $U$. Numerical evidence in \cite{lhl03} also supports the same statement.

The introduction of ancillae $R_A,R_B$ is necessary for both definitions of $K_E$ and $K_{Ea}$. For example, the SWAP gate on two qubits cannot create any entanglement starting from a pure state on $AB$; however, one can easily show that $K_E({\rm SWAP})=2$ ebits.
When the ancillae are not allowed, denote the restricted versions of $K_E$ and $K_{Ea}$ as ${\bar K}_E$ and ${\bar K}_{Ea}$, respectively. The paragraph after Eq.~(12b) of \cite{dvc01} implies that there is a $U$ such that ${\bar K}_{Ea}(U)>{\bar K}_E(U)$. This fact is also proved in \cite{lhl03}.

If the expression $E(U(\ket{\psi}))-E(\ket{\psi})$ is changed to $E(\ket{\psi})-E(U(\ket{\psi}))$ in \eqref{eq:K_ea}, the resulting quantity $K_d(U)$ is the so-called \textit{disentangling power} \cite{lsw09}. One can show that
\bea
\label{eq:kdu=keau}
K_d(U)=K_{Ea}(U^\dg),
\eea
and determine the properties of disentangling power via that of assisted entangling power. The disentangling power physically means the maximum entanglement that a bipartite unitary can annihilate. The disentangling power and assisted entangling power are generally different. In page 3 of \cite{lsw09}, a $2\times3$ non-controlled bipartite unitary $U$ has been constructed so that $K_{Ea}(U) = K_E(U) = 2> K_{Ea}(U^\dg)$. Since $K_{Ea}(U^\dg)\ge K_E(U^\dg)$, we have $K_E(U)>K_E(U^\dg)$.
It solves an open problem in \cite[Table 1]{Nielsen03}.

As the physical inverse of entangling power, we investigate the cost of creating bipartite unitaries. In this paper, the ``entanglement cost'' of a bipartite unitary $U$ is defined as
$
E_c(U)=\inf_p E_c(p),
$
where $p$ is any one-shot exact deterministic LOCC protocol for implementing $U$ with a pure entangled state as the nonlocal resource, and $E_c(p)$ is the amount of entanglement in the resource state, measured using the entanglement entropy.
The Schmidt rank of the pure state and the dimension of ancillary space are finite in the protocol $p$, and have no constant upper bound when taking the infimum.
%
We refer to $E_c(U)$ as the one-shot entanglement cost.
Define the asymptotic entanglement cost of a bipartite unitary $U$ as $E'_c(U):=\lim_{n\ra\infty} \frac{E_c(U^{\otimes n})}{n}$ and asymptotic assisted entangling power as $K'_{Ea}(U):=\lim_{n\ra\infty} \frac{K_{Ea}(U^{\otimes n})}{n}$. Since entanglement is non-increasing under LOCC, we have $E'_c(U)\ge K'_{Ea}(U)$. The definitions of the two types of entanglement costs and the definition of assisted entangling power
imply $K'_{Ea}(U)\ge K_{Ea}(U)$ and $E'_c(U)\le E_c(U)$. We have
\bea\label{eq:ecbound}
K_E(U)\le K_{Ea}(U)\le K'_{Ea}(U) \le  E'_c(U)\le E_c(U).
\eea
Hence, if $K_E(U)=E_c(U)$ then all quantities become  the same. This is exactly the case of generalized Clifford operators we investigate in Proposition \ref{prop:Clifford}. The question stated near the beginning of the introduction can be formalized as the following question: Is there a bipartite unitary $U$, such that $K_{Ea}(U) < E_c(U)$?

The rest of this paper is organized as follows.  In Sec. \ref{sec:pre} we introduce the notations and known results used  in the paper. In Sec. \ref{sec:ep} we investigate the entangling power of bipartite unitaries of Schmidt rank two, bipartite permutation unitaries, and $2\times d_B$ complex permutation matrices of Schmidt rank three. We also investigate non-controlled bipartite unitaries including Schmidt-rank-four $2\times d_B$ complex permutation unitaries and two-qubit unitaries. We further show the connection between our results and symmetric informationally complete positive operator-valued measure (SIC-POVM). In Sec. \ref{sec:aep} we investigate the assisted entangling power of bipartite unitaries. We derive the entangling power and assisted entangling power for generalized Clifford operators. We also present the concept of generalized CNOT gate. Such gates have the maximum entangling power  in arbitrary dimensions. In Sec. \ref{sec:eae} we study the relation between the entangling power, assisted entangling power and the disentangling power. In Sec. \ref{sec:open} we discuss two conjectures arising in the literature and this paper. We conclude in Sec. \ref{sec:con}.

\section{Preliminaries}\label{sec:pre}

In this section we introduce the notations and known results used in the paper. Denote the computational-basis states of the bipartite Hilbert space $\cH=\cH_A\ox\cH_B$ by $\ket{i,j},i=1,\cdots,d_A$, $j=1,\cdots,d_B$. Let
$I_A$ and $I_B$ be the identity
operators on the spaces $\cH_A$ and $\cH_B$, respectively. We also denote $I_d$ and $0_d$ respectively as the identity and zero matrix of order $d$. Any bipartite unitary gate $U$ acting on $\cH$ has \emph{Schmidt rank} [denoted as ${\rm Sch}(U)$] equal to $n$ if there is an expansion of the form $U=\sum^n_{j=1}A_j \ox B_j$ where the $d_A\times d_A$ matrices $A_1,\cdots,A_n$ are linearly independent, and the $d_B\times d_B$ matrices $B_1,\cdots,B_n$ are also linearly independent. The Schmidt rank is equivalent to the notion of operator-Schmidt rank in \cite{Nielsen03,Tyson03}. The above expansion is called the \emph{Schmidt decomposition}.
We can further write the Schmidt decomposition in a standard form,
\bea\label{eq:schmidtstandard}
U=\sum_{j=1}^r c_j A_j\ox B_j,
\eea
where $\frac{1}{d_A}\tr(A_j^\dag A_k)=\frac{1}{d_B}\tr(B_j^\dag B_k)=\delta_{jk}$, $c_j>0$, and $\sum_{j=1}^r c_j^2=1$. Then we introduce the \textit{Schmidt strength}
\bea\label{eq:schmidt_strength}
K_{Sch}(U)=-\sum_{j=1}^r c_j^2 \log_2 c_j^2
\eea
which is used as a measure of the ``nonlocal content'' of $U$ \cite{Nielsen03}.  The inequality
\bea
\label{eq:keu>=ksch}
\log_2 \sch (U) \ge K_E(U)\ge K_{Sch}(U)
\eea
holds for any bipartite unitary $U$ in terms of the definition of $K_E$ and \cite[Theorem 1]{Nielsen03}.

Next, $U$ is a \textit{controlled unitary gate}, if $U$ is equivalent to $\sum^{d_A}_{j=1}\proj{j}\ox U_j$ or
$\sum^{d_B}_{j=1}V_j \ox \proj{j}$ via local unitaries. To be specific, $U$ is a controlled unitary from $A$ or $B$ side, respectively.
In particular, $U$ is controlled in the computational basis from $A$ side if $U=\sum^{d_A}_{j=1}\proj{j}\ox U_j$. Bipartite unitary gates of Schmidt rank two or three are in fact controlled unitaries \cite{cy13,cy14,cy14ap}.
We shall denote $V\op W$ as the ordinary direct sum of two matrices $V$ and $W$, and $V\op_B W$ as the direct sum of $V$ and $W$ from the $B$ side (called ``$B$-direct sum''). In the latter case, $V$ and $W$ respectively act on two subspaces $\cH_A\ox\cH'_B$ and $\cH_A\ox\cH''_B$, respectively, such that $\cH_B'\perp\cH_B''$.
A permutation matrix (or called ``permutation unitary'') is a unitary matrix containing elements $0$ and $1$ only. A partial permutation matrix is obtained by changing some element $1$ to $0$ in a permutation matrix. A bipartite controlled-permutation matrix is a permutation matrix controlled in the computational basis of one system.
Each term in a controlled-permutation unitary refers to a term of the form $P\otimes V$ (or with the two sides swapped), where $P$ is a projector whose rank is a positive integer, and $V$ is a local permutation unitary. A ``big row'' of the $d_A d_B\times d_A d_B$ unitary matrix $U$ refers to a $d_B\times d_Ad_B$ submatrix given by $_A \bra{j}U$, for some $j\in\{1,\dots,d_A\}$. Similarly, a ``big column'' of $U$ refers to a $d_A d_B\times d_B$ submatrix given by $U\ket{j}_A$, for some $j\in\{1,\dots,d_A\}$. A ``block'' of $U$ refers to a $d_B\times d_B$ submatrix given by $_A \bra{j}U\ket{k}$, for some $j,k\in\{1,\dots,d_A\}$, and when $j=k$, the block is called a ``diagonal block.''

It is known that any controlled unitary controlled from the $A$ side on the space $\cH_A\ox\cH_B$ is locally equivalent to
\bea
\label{eq:mterms}
U=\sum^{m}_{j=1} P_j \ox U_j
\eea
where the $P_j$'s are pairwise orthogonal projectors on $\cH_A$, and the $U_j$'s are unitary operators on $\cH_B$. We can further assume that the $U_j$'s are pairwise linearly independent, and say that $U$ is controlled with $m$ terms.
Next we review mathematical results on von Neumann entropy, quantum channel, and controlled unitaries.


\subsection{Mathematics of quantum information}

Let $H(\{p_j\}) := \sum_j - p_j \log_2 p_j$ be the Shannon entropy of the probability distribution $\{p_j\}$. The following lemma (i) is known as the subadditivity of von Neumann entropy. It follows from the paragraph below (11.73) and Exercise 11.16 of \cite{nc2000book}. Lemma \ref{le:concave} (ii) is from (11.84) and Theorem 11.10 in \cite{nc2000book}. In particular the second inequality in \eqref{eq:concave} is known as the concavity of von Neumann entropy.

\bl
\label{le:concave}
(i) Let $\r_{AB}$ be a density operator on two systems $A,B$. Then
\bea
\abs{S(\r_A)-S(\r_B)} \le S(\r_{AB})\le S(\r_A)+S(\r_B).
\eea
The first equality holds if and only if there is a split of the system $A=A_1A_2$ such that $\r_{AB}=\proj{\ps}_{A_1B}\ox\s_{A_2}$, or there is a split of the system $B=B_1 B_2$ such that $\r_{AB}=\proj{\ps}_{A B_1}\ox\s_{B_2}$. The second equality holds if and only if $\r_{AB}=\r_A\ox\r_B$.
\\
(ii) Let $\{p_j\}$ be a probability distribution of $p_j>0$ and $\{\r_j\}$ a set of density operators. Then
\bea
\label{eq:concave}
\sum_j p_j S(\r_j) + H(\{p_j\})\ge S(\sum_j p_j \r_j) \ge \sum_j p_j S(\r_j).
\notag\\
\eea
The first equality holds if and only if the range of $\r_i$ and $\r_j$ are pairwise orthogonal, $\forall i,j$. The second equality holds if and only if $\r_i=\r_j$, $\forall i,j$.
\qed
\el
We will use this lemma to derive the entangling power of bipartite complex permutation unitaries of Schmidt rank three in Proposition \ref{pp:cp3}, and investigate the assisted entangling power of controlled unitaries in Proposition \ref{pp:aep,control}.
Below is a known result from the majorization theory.

\bl
\label{le:majorization}
Let $\r$ and $\s$ be two quantum states. If the spectrum of $\r$ is strictly majorized by the spectrum of $\s$, i.e., $\r\prec_s \s$, then $S(\r)> S(\s)$.
\el
We will use the lemma to derive the upper bound of entangling power of
a family of bipartite unitary operator of Schmidt rank three in Lemma \ref{le:nqp=220}. Let $\ket{\ps}=\sum^r_{j=1} \sqrt{p_j}\ket{a_j,b_j}$ be the Schmidt decomposition with nonnegative real numbers $p_j$ in the descending order. We refer to the vector $\sv(\ps)$ of probability distribution $(p_1,\cdots,p_r)$ as the \textit{Schmidt vector} of $\ket{\ps}$. For an arbitrary vector $x$ of probability distribution, we refer to $\des(x)$ as the vector whose elements are the same as those of $x$ except that they are in the descending order.
Next we show conditions by which a quantum channel converts an arbitrary input into the maximally mixed state.

\bl
 \label{le:unital}
For $d^2$ operators $K_1,\cdots,K_{d^2}$ and an invertible operator $R$ acting on $\bC^d$, the following five assertions are equivalent:
\\
(i) $\tr K_i^\dg R^{-1} K_j = \d_{ij}$ for $i,j=1,\cdots,d^2$;
\\
(ii) $\sum^{d^2}_{j=1} K_j^\dg X K_j = (\tr RX) I_d$ for all matrices $X$ acting on $\bC^d$;
\\
(iii) $\sum^{d^2}_{j=1} K_j^\dg X K_j = (\tr RX) I_d$ for all pure states $X$ acting on $\bC^d$;
\\
(iv) $\tr_A (\sum^{d^2}_{j=1} \proj{j}\ox K_j^\dg) Y (\sum^{d^2}_{j=1} \proj{j}\ox K_j) = \tr (R \cdot \tr_A Y) I_d$ for all bipartite operators $Y$ acting on $\cH=\bC^{d^2}\ox \bC^d$;
\\
(v) $\tr_A (\sum^{d^2}_{j=1} \proj{j}\ox K_j^\dg) Y (\sum^{d^2}_{j=1} \proj{j}\ox K_j) = \tr (R \cdot \tr_A Y) I_d$ for all pure product states $Y$ acting on $\cH=\bC^{d^2}\ox \bC^d$.
\el
\bpf
The equivalence between assertions (i) and (ii) is from \cite[Proposition 3]{werner01}. Assertion (iii) is equivalent to (ii) because the equation $\sum^{d^2}_{j=1} K_j^\dg X K_j = (\tr RX) I_d$ is linear with $X$, and any matrix space is spanned by rank-one positive semidefinite matrices. The same reason implies the equivalence between (iv) and (v). Finally, (ii) and (iv) are equivalent by setting $X=\tr_A Y$.
This completes the proof.
\epf

An important case of this lemma is when $R=I_d$.

\bcr
\label{cr:unital}
For $d^2$ operators $K_1,\cdots,K_{d^2}$ acting on $\bC^d$, the following five assertions are equivalent:
\\
(i) $\tr K_i^\dg K_j = \d_{ij}$ for $i,j=1,\cdots,d^2$;
\\
(ii) $\sum^{d^2}_{j=1} K_j^\dg X K_j = (\tr X) I_d$ for all matrices $X$ acting on $\bC^d$;
\\
(iii) $\sum^{d^2}_{j=1} K_j^\dg X K_j = I_d$ for all pure states $X$ acting on $\bC^d$;
\\
(iv) $\tr_A (\sum^{d^2}_{j=1} \proj{j}\ox K_j^\dg) Y (\sum^{d^2}_{j=1} \proj{j}\ox K_j) = (\tr Y) I_d$ for all bipartite operators $Y$ acting on $\cH=\bC^{d^2}\ox \bC^d$;
\\
(v) $\tr_A (\sum^{d^2}_{j=1} \proj{j}\ox K_j^\dg) Y (\sum^{d^2}_{j=1} \proj{j}\ox K_j) =I_d$ for all pure product states $Y$ acting on $\cH=\bC^{d^2}\ox \bC^d$.
\ecr
The corollary is used in the following discussion.
Assertion (i) implies that the set $\{K_j\}_{j=1,\cdots,d^2}$ is an orthonormal basis of the $d\times d$ matrix space
under the Hilbert-Schmidt inner product $\norm{A,B}_{hs}:=\tr(A^\dg B)$.
It occurs e.g., when $\{{K_j \over \sqrt d }\}_{j=1,\cdots,d^2}$ is the Heisenberg-Weyl (HW) group. In this case, assertion (ii) implies \cite[Exercise 11.19]{nc2000book}. Furthermore, assertion (ii) implies that the map $\L(\cdot):={1\over d^3}\sum^{d^2}_{j=1} K_j^\dg (\cdot) K_j$ is a depolarized channel and at the same time a unital channel because of $\L(I)=I$ \cite{king02}. The unital channels have been extensively studied in the past years \cite{dr05,mw09,fg15}.
In particular, the unitaries $K_j$ have been used to construct
mutually unbiased unitaries \cite{sm16}.
The following result is implied by \cite{br03}. See more general cases in \cite{Ambainis00,ns07}.

\bl
\label{le:da>=db2}
Let $U=\sum^{d_A}_{j=1} \proj{j} \ox U_j$ be a controlled unitary such that there is a constant state $\ket{\a}$ satisfying that for any state $\ket{\b}$, $U \ket{\a}_A\ket{\b}_B$ is maximally entangled. Then $d_A\ge d_B^2$.
\el
Since $d_A\ge d_B^2\ge d_B$,  $U \ket{\a}_A\ket{\b}_B$ is locally equivalent to the $d_B\times d_B$ maximally entangled state.
The condition of this lemma is equivalent to the statement that there is a constant state $\ket{\a}=\sum^{d_A}_{j=1} \sqrt{p_j}\ket{j}$, such that
$
\sum^{d_A}_{j=1} p_j U_j \proj{\b} U_j^\dg = {1\over d_B} I_B.
$
If $d_A=d_B^2$, then this equation is a special case of Corollary \ref{cr:unital} (iii). Since it is equivalent to Corollary \ref{cr:unital} (i), we can work out that $p_j={1\over d_B^2}$ for any $j$. Hence, $\{U_j\}$ must be a set of orthogonal unitary bases under the Hilbert-Schmidt inner product.
The following fact is mentioned in the paragraph of \cite[Eq. (15)]{rpl16}.
\bl
\label{le:diagonal}
For any $d\times d$ matrix $X$, the matrix ${1\over r}\sum^{r-1}_{k=0}U_k X U_k^\dg $ is diagonal when either of the following two conditions is satisfied:
\\
(i) $r=d$ and $U_k=\diag(1,\o^k,\cdots,\o^{k(r-1)})$ and $\o=e^{2\p i/d}$;
\\
(ii) $r=2^d$ and $U_k=\diag(\pm1,\pm1,\cdots,\pm1)$.
\el
The above two lemmas will be used to characterize the entangling power of bipartite unitaries below Lemma \ref{le:control}.
If either condition holds, then one can find out $d$ permutation matrices
$P_k:=\sum_{j=1}^d \ketbra{j}{1+(j+k-1)\mod d}$ for $k=1,\cdots,d$ and $Y_i:=P_{i}({1\over r}\sum^{r-1}_{k=0}U_k X U_k^\dg) P_{i}^\dg$. Then $\sum^{d}_{i=1} Y_i = (\tr X) I_d$, i.e., any matrix $X$ can be converted to the maximally mixed state under the unital channel. If the condition is (i), then one can verify that the set $\{{1\over\sqrt d}P_i U_k\}$ satisfies Corollary \ref{cr:unital} (i). So the set is a constructive example of the operators in Corollary \ref{cr:unital}. On the other hand, if the condition is (ii) then the set does not satisfy Corollary \ref{cr:unital} (i).

Finally we present a lemma for the block-controlled unitary (BCU) operations \cite{cy14}. The latter is defined as the direct sum of two bipartite unitaries from the $A$ or $B$ side (allowing the freedom of local unitaries). So a controlled unitary is a BCU and the inverse is wrong. The BCU is the generalization of the notion of controlled unitaries. The Lemma~\ref{le:bcu} below will be used to show that any bipartite permutation unitary of Schmidt rank greater than two has entangling power greater than $1.223$ ebits; see Proposition~\ref{pp:>1ebit}. We also define the \emph{block-controlled-permutation unitary} (BCPU) as a BCU which is block diagonal in the standard basis on the controlling side and is at the same time a permutation unitary in the standard basis. This notion will be used in the proof of Proposition~\ref{pp:>1ebit}.

\bl
\label{le:bcu}
Let $U=V\op_B W$ be a bipartite unitary. Then $K_E(U)\ge \max\{K_E(V),K_E(W)\}$.
\el
\bpf
By the equation $U=V\op_B W$, we have $\cH_B={\cH_B^V}\oplus {\cH_B^W}$, where the subspace ${\cH_B^V}$ (respectively, ${\cH_B^W}$) is the input subspace of $V$ (respectively, $W$). Denote the input state on $B R_B$ as $\ket{\phi}_{B R_B}$.
The inequality follows by restricting the reduced density matrix $\tr_{R_B}(\proj{\phi}_{B R_B})$ to have support in the subspaces ${\cH_B^V}$ and ${\cH_B^W}$, respectively.
This completes the proof.
\epf

\section{Entangling power of bipartite unitaries}
\label{sec:ep}

Two main classes of bipartite unitary operations are bipartite controlled unitaries and permutation unitaries. The former contains the basics of quantum circuits, such as CNOT gates and controlled-phase gates. Next, any bipartite unitary is the product of controlled unitaries \cite{bry02,blb05}. Any bipartite controlled unitary can be implemented with LOCC and a maximally entangled state \cite{ygc10}, thus a general bipartite unitary can be implemented by performing the controlled unitaries in its decomposition. The implementation is more efficient for bipartite unitaries of Schmidt rank at most three, because they are equivalent to controlled unitaries under local unitaries \cite{cy13,cy14,cy14ap}. In particular, any bipartite permutation unitary of Schmidt rank three can be implemented using LOCC and two ebits \cite{cy16}. On the other hand, a protocol for implementing bipartite permutation unitaries of any Schmidt rank $r$ has been given, by using $O(r)$ ebits of entanglement and $O(r)$ bits of classical communication \cite{cy16}. These facts imply that the two classes of bipartite unitaries are experimentally available resources. So the next step is to understand their entangling power in practice.

We begin by studying the entangling power of bipartite controlled unitaries in Lemma \ref{le:control}, and then apply it to some well-known bipartite unitaries in subsections. The latter includes Schmidt-rank-two unitaries in Sec. \ref{subsec:sr2}, Schmidt-rank-three permutation unitaries and $2\times d_B$ complex permutation matrices in Sec. \ref{subsec:sr3permutation}, and non-controlled bipartite unitaries such as a family of bipartite unitaries including the SWAP gate as a proper subset, and Schmidt-rank-four $2\times d_B$ complex permutation unitaries in Sec. \ref{subsec:non}. We further show that any bipartite permutation unitary of Schmidt rank greater than two has entangling power greater than $1.223$ ebits in Proposition \ref{pp:>1ebit}. We also point out the connection between the controlled unitaries and the symmetric informationally complete positive operator-valued measure (SIC-POVM) in Sec. \ref{sub:sic}.

If $\ket{\a,\b}$ maximizes $E(U(\ket{\alpha}\ket{\beta}))$ in \eqref{eq:K_e}, then we call it the critical state of $U$. In general, a bipartite unitary has many critical states. The critical states of bipartite controlled unitaries have a simpler structure, as we show below.


\bl
\label{le:control}
Suppose $U=\sum^{m}_{j=1} P_j \ox U_j$ in \eqref{eq:mterms} is a controlled unitary controlled with $m$ terms. Then
(i)
\bea
\label{eq:K_eu}
K_E(U)
&=&
\max_{\ket{\alpha}\in\cH_A,\ket{\beta}\in\cH_{BR_B}} E(U(\ket{\alpha}\ket{\beta}))
\notag\\
&=&\max_{p_j\ge0,~\sum^{m}_{j=1} p_j=1,~\ket{\beta}\in\cH_{BR_B}}
\notag\\
&&S
\bigg(
\sum^{m}_{j=1} p_j (U_j)_B \proj{\b}_{BR_B} (U_j)_B^\dg
\bigg)
\notag\\
&\le& \log_2 \sch(U)
\notag\\
&\le& \log_2 \min\{m,d_B^2\}.
\eea
In particular, $K_E(U)=\log_2 \sch(U)$ if and only if
$
\sum^{m}_{j=1} p_j (U_j)_B \proj{\b}_{BR_B} (U_j)_B^\dg$ is a normalized projector of rank $\sch(U)$.
\\
(ii) If $U$ is also controlled from the $B$ side, then
\bea
\label{eq:K_euab}
K_E(U)
&=&
\max_{\ket{\alpha}\in\cH_A,\ket{\beta}\in\cH_{B}} E(U(\ket{\alpha}\ket{\beta}))
\notag\\
&=&\max_{p_j\ge0,~\sum^{m}_{j=1} p_j=1,~\ket{\beta}\in\cH_{B}}
\notag\\
&&S
\bigg(
\sum^{m}_{j=1} p_j (U_j)_B \proj{\b}_{B} (U_j)_B^\dg
\bigg)
\notag\\
&\le& \log_2 \sch(U).
\eea
In particular, $K_E(U)=\log_2 \sch(U)$ if and only if $
\sum^{m}_{j=1} p_j (U_j)_B \proj{\b}_{B} (U_j)_B^\dg$ is a normalized projector of rank $\sch(U)$.
\\
(iii) If $U$ is not controlled from the $B$ side, then
\bea\label{eq:K_euab2}
&&
K_E(U)\ge
\max_{\ket{\alpha}\in\cH_A,\ket{\beta}\in\cH_{B}} E(U(\ket{\alpha}\ket{\beta})),
\eea
and the inequality may hold or not.
\\
(iv) Let $\ket{\a,\b}$ be the critical state of $U$. Then $\ket{\a}$ can be chosen as a linear combination of the computational basis states with non-negative coefficients. If all $U_j$ are diagonal, then $\ket{\b}\in\cH_B$ can be chosen to also possess the same property.
\el
The proof is given in Appendix \ref{app:{le:control}}. Assertion (i) implies that the ancilla in the controlling side of a controlled unitary cannot increase the entangling power of the unitary. Note that an upper bound of the entanglement cost of controlled unitary from the $A$ side with $d_A=2,3$ is $\log_2 \min\{d_A^2, d_B\}$ \cite{cy14}. It is similar to that in (i), which is an upper bound of the entangling power. The entangling power is upper bounded by the entanglement cost with two upper bounds, namely $\log_2 \min\{d_A^2, d_B\}$ and $\min \{
\log_2 \sch(U),
~
\log_2 d_A,
~
2\log_2 d_B
\}$.
On the other hand, the trivial upper bound $K_E(U)\le\log_2 \sch(U)$ is again obtained in spite of the simplification by the controlled unitaries. A tighter upper bound might be achievable only if the considered controlled unitaries are restricted to a smaller subset of controlled unitaries.

Next, assertion (ii) implies that the unitary is controlled from both sides; then we can discard both ancillae in \eqref{eq:K_e}. For example, the critical state of a Schmidt-rank-two unitary \cite{cy13}, or a Schmidt-rank-three diagonal unitary need not include any ancilla system.
On the other hand, for controlled unitaries $U$ whose $B$ side cannot be the controlling system, the ancilla system $R_B$ in \eqref{eq:K_eu} cannot generally be removed because of (iii).

The first example in (iii) is not a permutation matrix. Here we give an example of permutation matrix. Let $V=\sum^4_{j=1}\proj{j}\ox P_j$, where $P_1=I_3$, $P_2=\proj{1}+\ketbra{2}{3}+\ketbra{3}{2}$, $P_3=\proj{2}+\ketbra{3}{1}+\ketbra{1}{3}$, and $P_4=\proj{3}+\ketbra{1}{2}+\ketbra{2}{1}$ act on the space $\cH_{AB}$. So $V$ is a bipartite permutation matrix. One can show by calculation that $V(\frac{1}{2\sqrt{3}}\sum^4_{j=1}\ket{j}_A \ox \sum^3_{k=1}\ket{kk}_{BB_R})$ has entanglement more than $\log_2 3$ ebits, which is the upper bound of the entangling power of $V$ without an ancilla. Hence,
the inequality in Eq.~\eqref{eq:K_euab2} holds for $V$.

Suppose $U$ in Lemma \ref{le:da>=db2} is also controlled from the $B$ side. If the ``constant'' in Lemma \ref{le:da>=db2} is removed, then the condition of this lemma means that $S
\bigg(
\sum^{d_A}_{j=1} p_j (U_j)_B \proj{\b}_{B} (U_j)_B^\dg
\bigg)= \log_2 \min\{d_A,d_B\}$. So the upper bound in \eqref{eq:K_euab} is saturated, and the equation $d_A\ge d_B^2$ might no longer hold. On the other hand, we do not know the case when $U$ in Lemma \ref{le:da>=db2} is not controlled from the $B$ side.

If the unitaries $U_i$ in \eqref{eq:K_euab} are those in either case of Lemma \ref{le:diagonal}, then we can work out that $K_E(U)=\log d_B$. In the following subsections, we investigate several types of bipartite unitaries and analytically derive their entangling power using Lemma \ref{le:control}.

\subsection{Schmidt-rank-two unitaries}
\label{subsec:sr2}

In this subsection we provide the analytical method of computing the entangling power of Schmidt-rank-two bipartite unitaries $U$.
It is known \cite{cy13} that up to local unitaries, $U$ is a controlled unitary and can be written as the form
\bea
\label{eq:up1}
U=P\ox I_{d_B}+(I_A-P)\ox D
\eea
where $P$ is a projector, and $D=\diag(e^{i\t_1},\cdots,e^{i\t_{d_B}})$ is a diagonal unitary with real $\t_1,\cdots,\t_{d_B}\in[0,2\p)$ in the ascending order.
It suffices to work with $U$ in the above form because the entangling power is invariant up to local unitaries. Lemma \ref{le:control} (ii) implies that
\bea
\label{eq:sr2complex}
K_E(U)
&=&\max_{p\in[0,1],~~~\ket{\beta}=(b_1,\cdots,b_{d_B})^T,~~~b_j\ge0}
\notag\\
&&S
\bigg(
p\proj{\b} + (1-p) D\proj{\b} D^\dg
\bigg),
\eea
where the components $b_j\ge0$ follow from the fact that the von Neumann entropy is invariant up to unitaries. Let $V$ be a $d_B \times d_B$ unitary whose first row is $(b_1,\cdots,b_{d_B})$. Applying the same fact to \eqref{eq:sr2complex} we obtain
\bea
\label{eq:sr2complex2}
K_E(U)
&=&
\max_{p\in[0,1],~~~\ket{\beta}=(b_1,\cdots,b_{d_B})^T,~~~b_j\ge0}
\notag\\
&&S
\bigg(
pV\proj{\b}V^\dg + (1-p) VD\proj{\b} D^\dg V^\dg
\bigg)
\notag\\
&=&
\max_{p\in[0,1],~~~\ket{\b'}=(x,\sqrt{1-x^2})^T,~~~x=\abs{\sum_j e^{i\t_j} b_j^2}
\atop b_j\ge0,~~~\sum_j b_j^2 = 1}
\notag\\
&&S
\bigg(
p\proj{0} + (1-p) \proj{\b'}
\bigg).
\eea
The maximum is achievable if and only if the determinant of the $2\times2$ matrix in the last row of \eqref{eq:sr2complex2} reaches the maximum. It implies $p=1/2$. Using \eqref{eq:sr2complex} we have
\bea
\label{eq:keu-sr2complex}
&&
K_E(U)=\max_{b_1,\cdots,b_{d_B},~~~\sum_j b_j^2 = 1}
\notag\\
&&
H\bigg( {1-\abs{\sum_j e^{i\t_j} b_j^2}\over2},{1+\abs{\sum_j e^{i\t_j} b_j^2}\over2} \bigg).
\eea
Setting $c_j=b_j^2$, we have $\sum_j c_j=1$. Hence,
\bea
\abs{\sum_j e^{i\t_j} b_j^2}
&=&
\bigg[ (\sum_j c_j\cos\t_j)^2 + (\sum_j c_j\sin\t_j)^2  \bigg]^{1\over2}
\notag\\
&=&
\bigg[ \sum_j c_j^2 + 2\sum_{j> k} c_j c_k\cos(\t_j-\t_k)  \bigg]^{1\over2}
\notag\\
&=&
\bigg[ 1 - 2 \sum_{j>k} c_j c_k + 2\sum_{j>k} c_j c_k\cos(\t_j-\t_k)  \bigg]^{1\over2}
\notag\\
&=&
\bigg[ 1 - 4 \sum_{j>k} c_j c_k\sin^2({\t_j-\t_k\over2})  \bigg]^{1\over2}.
\eea
So the minimum of $\abs{\sum_j e^{i\t_j} b_j^2}$, equivalently $K_E(U)$ in \eqref{eq:keu-sr2complex}, is reached at the maximum of
$
y(\{c_j\}):=\sum_{j>k} c_j c_k\sin^2({\t_j-\t_k\over2}),
$
where the parameters $c_j\ge0$ and $\sum_j c_j=1$.
If $d_B=2$, then straightforward computation shows that $K_E(U)$ in \eqref{eq:keu-sr2complex} is reached when $c_1=c_2={1\over 2}$. We have the following.
\bl
\label{le:keu-sr2complex}
Any $d_A\times2$ controlled unitary $U=P_1\ox I_{2}+P_2\ox \diag(e^{i\t_1},e^{i\t_2})$ with orthogonal projectors $P_1,P_2$ and the real parameter $\t$ has
the entangling power $K_E(U)=H \big( {1-\abs{\cos{\t_1-\t_2\over2}}\over2},{1+\abs{\cos{\t_1-\t_2\over2}}\over2} \big)$.
\el
If $d_A=2$, then the lemma reduces to the result in \cite[Theorem 2]{Nielsen03}.
In particular, the entangling power of two-qubit controlled unitaries is the same as the Schmidt strength in terms of Theorem 2 of \cite{Nielsen03}. Lemma \ref{le:keu-sr2complex} thus provides the analytical formula for the Schmidt strength of two-qubit controlled unitaries.
On the other hand, Lemma \ref{le:keu-sr2complex} implies that $K_E(U)$ reaches the maximum 1 ebit if and only if $\t_1-\t_2=(2k+1)\p$ for $k\in\bZ$. When $d_A=2$ the gate $U$ is locally equivalent to the CNOT gate. Besides, \eqref{eq:keu-sr2complex} for $d_B=2$ also generalizes the result in \cite{Nielsen03}.

Next, if $d_B>2$, then we use the equations ${\partial \big( y(\{c_j\}) + \l (\sum_j c_j-1)\big) \over \partial c_j}=0$ where $\l$ is the Lagrange multiplier. One can show that at most two of these equations are independent. So we have $\l=-1/2$, and thus
$
\sum_j c_j \sin^2({\t_1-\t_j\over2}) = {1\over2},
\sum_j c_j \sin^2({\t_2-\t_j\over2}) = {1\over2},
$ and $
\sum_j c_j  = 1.$
For given $\t_j$ we can derive the set of roots $c_j$ of the above linear equations. On the other hand, we need to study the boundary case. By setting some $c_j=0$  in \eqref{eq:keu-sr2complex} we can similarly obtain the above equations and work out the remaining $c_j$. They give rise to another set of roots $c_j$. Repeating this procedure, we obtain a few different sets of roots $c_j$. We input these sets in the binary function in \eqref{eq:keu-sr2complex} and obtain corresponding output values. The maximum of these values is equal to $K_E(U)$. So we can analytically work out $K_E(U)$. For example, using the above arguments and $h(i,j):=H \big( {1-\abs{\cos{\t_i-\t_j\over2}}\over2},{1+\abs{\cos{\t_i-\t_j\over2}}\over2} \big)$ we can derive the entangling power of $U$ with $d_B=3$.
\bpp
\label{pp:db=3}
Let $U$ be \eqref{eq:up1} with $d_B=3$.
We have $K_E(U)=\max \{h(1,2),h(2,3),h(1,3)\}$.
\epp

This result and Lemma \ref{le:keu-sr2complex} show the following conjecture for $n=2,3$.
\bcj
\label{cj:sr2}
For the Schmidt-rank-two bipartite unitary $V=\proj{1}\ox I_n + \proj{2} \ox \sum^n_{j=1} e^{i\t_j}\proj{j}$, we have
$
K_E(V)=\max_{1\le i<j\le d_B} \{h(i,j)\}.
$
\ecj

Using the results in this subsection, we further study the maximum of entangling and assisted entangling power of Schmidt-rank-two bipartite unitaries, namely the generalized CNOT gates in Sec. \ref{subsec:gcnot}.

\subsection{Schmidt-rank-three permutation unitaries}
\label{subsec:sr3permutation}

Finding the entangling power of an arbitrary Schmidt-rank-three bipartite unitary is a technically involved problem. We investigate the permutation operations. They are controlled unitaries \cite{cy14ap} though are not always controlled permutation unitaries. The main result is presented in Proposition \ref{pp:entpower3} and was proposed as an open problem in \cite{cy16}. We further derive the entangling power of Schmidt-rank-three $2\times d_B$ permutation operations in Proposition \ref{pp:cp3}. First we present a preliminary lemma proved in Appendix \ref{app:{le:nqp=220}}.
\bl
\label{le:nqp=220}
Consider the bipartite unitary operator of Schmidt rank three,
\bea\label{eq:perm_u_3terms}
U
&=& D_1\ox I_B
\notag\\
&+& D_2 \ox (I_m \op I_n \op V_1)
\notag\\
&+& D_3 \ox (I_m \op V_3 \op I_q),
\eea
where $D_j$ are nonzero and satisfy $D_jD_k=\d_{jk} D_j$, $\sum_j D_j = I_A$, and $V_1$ and $V_3$ are respectively of size $q\times q$ and $n\times n$. Then $K_E(U)\le\log_2 9 - 16/9$ ebits. The equality is saturated when $U$ is a permutation unitary.
\el
The considered $U$ is a special case of the bipartite unitaries
\bea
\label{eq:ud1}
&&
D_1\ox I_B
\notag\\
&+& D_2 \ox (I_m \op I_n \op V_1 \op V_2)
\notag\\
&+& D_3 \ox (I_m \op V_3 \op I_q \op V_4),
\eea
where $D_j$ are nonzero and $D_jD_k=\d_{jk} D_j$, $\sum_j D_j = I_A$, and $V_1,V_2,V_3$, and $V_4$ are permutation matrices. $V_1$ and $V_3$ are, respectively, of size $q\times q$ and $n\times n$, and both $V_2$ and $V_4$ are of size $p\times p$ where $p=d_B-m-n-q$. If $V_1$ or $V_3$ contains a nonzero diagonal entry, then we can move the entry by local permutation matrices on $\cH_B$ so that $I_m$ is replaced with $I_{m+1}$. So $V_1$ and $V_3$ do not contain any nonzero diagonal entry. Similarly, we may assume that $V_2$ and $V_4$ do not have a nonzero diagonal entry in the same column when $p>0$. Now we state the main result of this subsection.
\bpp
\label{pp:entpower3}
The entangling power of any bipartite permutation unitary of Schmidt rank three can only take one of two values: $\log_2 9 - 16/9$ or $\log_2 3$ ebits. The former occurs if and only if the unitary is of the form of \eqref{eq:ud1} and $p=0$.
\epp
\bpf
Let $U$ be the bipartite permutation unitary of Schmidt rank three in the assertion. It was shown in the proof of \cite[Proposition 1]{cy16} that if $U$ is not of the form \eqref{eq:ud1}, then the entangling power of $U$ is exactly $\log_2 3$ ebits. The same conclusion holds when $U$ is of the form of \eqref{eq:ud1} and $p>0$.
It remains to prove the assertion when $U$ is of the form of \eqref{eq:ud1} and $p=0$. This is a special case of Lemma~\ref{le:nqp=220} where $U$ is a permutation unitary; thus, $K_E(U)=\log_2 9 - 16/9$ ebits.
This completes the proof.
\epf
To generalize this result, we derive $K_E(U)$ when $U$ is a complex permutation unitary on the $2\times d_B$ system. We present the following lemma, which is clear.
\bl
\label{le:dax2=sr3}
Let $U=\sum^2_{j,k=1} \ketbra{j}{k}\ox U_{jk}$ be a $2\times d_B$ complex permutation unitary of Schmidt rank three. Then either $U_{11}\propto U_{22}$ or $U_{12}\propto U_{21}$.
\el
Up to local permutation unitaries, we may assume that $U_{11}=U_{22}=I_n\op 0_{d_B-n}$, $U_{12}=0_n\op I_{d_B-n}$, and $U_{21}=0_n\op C$ where $C$ is a complex permutation unitary of order $d_B-n$. Let the entangling power of $\ketbra{1}{2}\ox I_B + \ketbra{2}{1} \ox C$ be $M$. This leads us to the following proposition.
\bpp
\label{pp:cp3}
\bea
\label{eq:cp3}
K_E(U) = H({1\over e^M+1},{e^M \over e^M+1})+{e^M \over e^M+1}M.
\eea
\epp
The proof is given in Appendix \ref{app:{pp:cp3}}. By computation we can show that $K_E(U)$ monotonically increases with $M$. So $K_E(U)$ reaches its lower and upper bound, respectively, at $M=0$ and $M=1$, hence $1\le K_E(U)\le 1.57100011... < \log_2 3 \approx 1.585$ (ebits). So any $2\times d_B$ complex permutation unitary of Schmidt rank three cannot reach the maximum. We qualitatively explain this result as follows. Let us first consider the case that the initial state on $B R_B$ is a maximally entangled state. When the initial state on $A R_A$ is a maximally entangled state,
there are two terms among the four terms in the output state that are proportional to each other on the $B R_B$ side, e.g., the terms corresponding to $U_{11}$ and $U_{22}$ in the case $U_{11}\propto U_{22}$,
so in the Schmidt decomposition of the output state, the three terms are not of equal weight;
hence, the entangling power is less than $\log_2 3$ ebits.
The case of other initial states on $A R_A$ are also similar because the two terms from $V$ are of less weight than the remaining term. Finally, the case of other initial states on $B R_B$ is also similar.

We remark that since there are only two $2\times2$ permutation unitary matrices $I_2$ and $\s_x$, the $d_A\times2$ controlled-permutation unitary has Schmidt rank of at most two.
So any $d_A\times2$ Schmidt-rank-three bipartite unitary, which may be a permutation unitary, is not locally equivalent to a controlled-permutation unitary.

\subsection{Non-controlled bipartite unitaries}
\label{subsec:non}

In previous subsections we have investigated the entangling power of Schmidt-rank-two bipartite unitaries and Schmidt-rank-three bipartite permutation matrices. They are both controlled unitaries, while we often deal with more non-controlled unitaries in practice.
This is a harder problem and we investigate four examples. In the first example, we compute the entangling power of a special bipartite unitary in Proposition \ref{pp:gs}. The unitary includes the bipartite SWAP gate of arbitrary dimensions. Next we show that any $2\times d_B$ complex permutation unitary of Schmidt rank four has entangling power 2 ebits in Proposition \ref{pp:dax2}. We further show in Proposition \ref{pp:>1ebit} that any bipartite permutation unitary of Schmidt rank greater than two has entangling power greater than $1.223$ ebits. Finally, we investigate two-qubit unitaries.

The first example is a family of bipartite unitaries on $d_A \times d_B$ space with $d_A \le d_B$,
\bea
\label{eq:gs}
U_{AB}=\sum^{d_A}_{j,k=1} \ketbra{j}{k}\otimes V_{jk},
\eea
where the $d_B\times d_B$ submatrices $V_{jk}$ satisfy the following two properties: (1) for any given $k$, the value $\tr V_{jk}^\dg V_{jk}$ is either zero or constant $c_k$, and (2) if we put the entries of them in the same $d_B \times d_B$ matrix, then any two entries are in different positions of the matrix. Suppose $V_{jk}$ and $V_{j'k'}$ are nonzero blocks and $\ket{\ps}={1\over\sqrt{d_B}}\sum^{d_B}_{i=1}\ket{i}_B \ket{i}_{R_B}$ is a maximally entangled state of Schmidt rank $d_B$ on $B R_B$, where $R_B$ is an ancillary system.
The properties imply that the two non-normalized states $V_{jk}\ket{\ps}$ and $V_{j'k'}\ket{\ps}$ have the same modulus and
are orthogonal. If we perform $U_{AB}$ on the non-normalized input state $(\sum_k {1\over \sqrt{c_k}} \ket{kk})_{AR_A}\ket{\ps}_{BR_B}$, then we obtain a maximally entangled state of Schmidt rank $\sch (U_{AB})$. Thus $K_E(U)\ge \log_2 \sch (U_{AB})$ (ebits). On the other hand, by Lemma \ref{le:control} we have $K_E(U)\le \log_2 \sch (U_{AB})$. We conclude the above argument as follows.
\bpp
\label{pp:gs}
For unitaries of the form \eqref{eq:gs} we have $K_E(U_{AB})=\log_2 \sch (U_{AB})$.
\epp
Note that \eqref{eq:gs} may be not a complex permutation matrix. An example is the $2\times3$ bipartite unitary
\bea
U_{AB}=
\left[
           \begin{array}{cccccc}
             1/\sqrt2 & 0 & 0 & 0 & 0 & 1/\sqrt2\\
             0 & 1 & 0 & 0 & 0 & 0\\
             0 & 0 & 0 & 1 & 0 & 0\\
             0 & 0 & 0 & 0 & 1 & 0\\
             1/\sqrt2 & 0 & 0 & 0 & 0 & -1/\sqrt2\\
             0 & 0 & 1 & 0 & 0 & 0\\
           \end{array}
         \right].
\eea
The example satisfies that $c_k$ is constant for all $k$. In this case it is easy to verify that $K_E(U_{AB}^\dg)=K_E(U_{AB})=\log_2 \sch (U_{AB})$ ebits, and the input state is the same as before.


In the second example,
we investigate the entangling power of the $2\times d_B$ permutation unitary $U$.
If it has Schmidt rank two or three, then $K_E(U)$ has been respectively derived in Lemma \ref{le:keu-sr2complex} and Proposition \ref{pp:cp3}.
So it suffices to investigate the Schmidt-rank-four case.
\bpp
\label{pp:dax2}
Any $2\times d_B$ complex permutation unitary of Schmidt rank four has entangling power of 2 ebits.
\epp
\bpf
Let $U$ be a $2\times d_B$ complex permutation unitary of Schmidt rank four.
Up to local complex permutation unitaries, we may assume that $U=\sum^2_{j,k=1} \ketbra{j}{k}\ox U_{jk}$ where $U_{11}=I_k\op 0_{d_B-k}$, $0< k<d_B$, and the zero columns (if any) among the rightmost $d_B-k$ columns of $U_{12}$ are in the rightmost columns of $U_{12}$. If a zero column exists, then there are four integers $i\in[1,k]$, $j\in[k+1,d_B-k]$ and $m,n\in[1,d_B]$, $m\ne n$, such that $U$ contains four nonzero entries in the positions $\ketbra{1,i}{1,i}$, $\ketbra{1,j}{2,i}$, $\ketbra{2,m}{1,d_B}$, and $\ketbra{2,n}{2,d_B}$. By performing $U$ on the input state ${1\over\sqrt2}(\ket{11}+\ket{22})_{AR_A}{1\over\sqrt2}(\ket{i,i}+\ket{d_B,d_B})_{BR_B}$, we obtain a uniformly entangled state of Schmidt rank four. So the assertion holds. On the other hand, if the rightmost $d_B-k$ columns of $U_{12}$ do not contain a zero column, up to local complex permutation unitaries we can assume that $U_{12}=0_k\op 1_{d_B-k}$. Since $U$ has Schmidt rank four, there are five integers $i\in[1,k]$, $j\in[k+1,d_B-k]$, and $m,n,p\in[1,d_B]$, $p\ne i$ and $m\ne n$, such that $U$ contains four nonzero entries in the positions $\ketbra{1,i}{1,i}$, $\ketbra{1,m}{2,j}$, $\ketbra{2,n}{1,j}$, and $\ketbra{2,p}{2,i}$. By performing $U$ on the input state ${1\over\sqrt2}(\ket{11}+\ket{22})_{AR_A}{1\over\sqrt2}(\ket{ii}+\ket{jj})_{BR_B}$, we obtain a uniformly entangled state of Schmidt rank four. So the assertion holds.
This completes the proof.
\epf

We have investigated the entangling power of many sorts of permutation unitaries. Below, we investigate the lower bound of entangling power of all permutation unitaries. The proof is given in Appendix \ref{app:{pp:>1ebit}}.

\bpp
\label{pp:>1ebit}
Any bipartite permutation unitary of Schmidt rank greater than two has entangling power of greater than 1.223 ebits.
\epp
The result shows that the entangling power of permutation unitaries is generally greater than that of arbitrary bipartite unitaries, because the latter with any Schmidt rank could have entangling power close to zero. An example of such bipartite unitaries is the controlled unitary $\sum^{d_A}_{j=1}\proj{j}\ox U_j$, where the linearly independent $U_j$ are close to the identity matrix.

Finally, we estimate the entangling power of two-qubit unitaries. It is known \cite{Nielsen03} that any two-qubit unitary gate is locally equivalent to
$
U=c_0 I_2\ox I_2 + c_x\s_x\ox\s_x +c_y\s_y\ox\s_y +c_z\s_z\ox\s_z
$
with complex numbers $c_0,c_x,c_y$ and $c_z$. We perform $U$ on the product states
${1\over\sqrt2}(\ket{11}+\ket{22})_{AR_A}\ox\ket{\ps}_{BR_B}$,
where $\ket{\ps}_{BR_B}$ is an arbitrary two-qubit state.
The resulting state is locally equivalent to $\ket{\Ps}=\sum_{j=0,x,y,z} c_j\ket{a_j,\ps_j}$, where $\ket{a_j}$ is an orthonormal basis in $\bC^4$.
It follows from \cite[Corollary 4]{Nielsen00} that $\sv(\ps) \succ \des(\abs{c_0}^2,\abs{c_x}^2,\abs{c_y}^2,\abs{c_z}^2)$.
It follows from Lemma \ref{le:majorization} that  $H(\sv(\ps))\le H(\abs{c_0}^2,\abs{c_x}^2,\abs{c_y}^2,\abs{c_z}^2)$. The equality is achievable when $\ket{\ps}$ is the two-qubit maximally entangled state.
So we obtain
$
K_E(U)\ge\sum_{j=0,x,y,z} -\abs{c_j}^2 \log_2 \abs{c_j}^2.
$
This result is exactly \cite[Theorem 1]{Nielsen03}, and the right-hand side of this equation is equal to the Schmidt strength. The result also coincides with \eqref{eq:schmidt_strength}. It is believed that the strict inequality holds for some $U$.

\subsection{Connection with SIC-POVM}

\label{sub:sic}

In a $d$-dimensional Hilbert space, the SIC-POVM \cite{rbs04} consists of $d^2$ outcomes that are subnormalized projectors
onto pure states $\frac{1}{d}\proj{\ps_j}$ for
$j=1,\ldots,d^2$, such that
$
\abs{\braket{\ps_j}{\ps_k}}^2=\frac{1+d\delta_{jk}}{d+1}.
$
Hence
$
\sum^{d^2}_{j=1}\proj{\ps_j}=d I_d.$
Many known SIC-POVMs are generated by performing the Heisenberg-Weyl (HW) group $\{U_j\}_{j=1,\cdots,d^2}$ on the so-called fiducial state $\ket{\ph}$ such that
$\ket{\ps_j}=U_j\ket{\ph}$ and
$\abs{\bra{\ph}U_j\ket{\ph}}={1\over\sqrt {d+1}}$,
where $U_j\ne I_d$. In the following we relate the SIC-POVM to \eqref{eq:K_euab} in Lemma \ref{le:control}. If the $U_j$ in \eqref{eq:K_euab} form the HW group, $\ket{\b}$ in \eqref{eq:K_euab} is a fiducial vector, $d_A=d_B^2$, then we can set $p_j={1\over d^2}$ in \eqref{eq:K_euab} for all $j$ and obtain $K_E(U) =\log_2 d_B$ because of $\ket{\ps_j}=U_j\ket{\ph}$. Physically, it means that
the reduced density operator on $B$ for the output state of the $U$ in \eqref{eq:K_euab} can always be chosen to be the maximally mixed state for some suitable input state.
As far as we know, this is the first necessary condition of the existence of fiducial-state-generated SIC-POVM in terms of the entangling power of controlled unitaries. On the other hand, if the fiducial-state-generated SIC-POVM does not exist in some $\bC^d$, the last equality in \eqref{eq:K_euab} still holds when $d_A=d_B^2$, in terms of Corollary \ref{cr:unital}. So the above necessary condition may be not sufficient, though we do not know the existence of SIC-POVM.
An interesting question is whether the ``fiducial-state-generated'' can be removed from the above discussion. It is an open problem whether the existence of SIC-POVM in $\bC^d$ implies the existence of fiducial-state-generated SIC-POVM in $\bC^d$ \cite{privatezhu}, though the converse evidently holds.

\section{Assisted entangling power of bipartite unitaries}
\label{sec:aep}

In this section we investigate the assisted entangling power of bipartite unitaries.
By definition, the input states can be arbitrary pure states with reference systems. Hence, the derivation of assisted entangling power is a harder problem than that of entangling power. In Proposition \ref{pp:aep,control}, we construct the upper bound for the assisted entangling power of controlled unitaries, and the necessary and sufficient condition by which the bound is saturated. Further, we introduce two families of (non-controlled) bipartite unitaries: the generalized CNOT gates in  Sec. \ref{subsec:gcnot} and the generalized Clifford gates in Sec. \ref{subsec:clifford}. The GCNOT gate has the maximum entangling and assisted entangling power among Schmidt-rank-two bipartite unitaries of high dimensions. So the GCNOT gate plays the same role as the CNOT gate does in the two-qubit unitary gates. We will derive the entangling power and assisted entangling power of both gates in Propositions \ref{pp:gcnot} and \ref{prop:Clifford}, respectively. Further, the asymptotic entangling and assisted entangling power, and the disentangling power of Clifford gates are also derived in Proposition \ref{prop:Clifford}.


\bpp
\label{pp:aep,control}
Suppose $U=\sum^{m}_{j=1} P_j \ox U_j$ in \eqref{eq:mterms} is a controlled unitary controlled with $m$ terms. Then (i)
\bea
\label{eq:K_aep}
&&
\log_2 m
\notag\\
&\ge&
K_{Ea}(U)
\notag\\
&=&
\max_{\sum^{m}_{j=1} M_j  =\r\in\cS(\cH_{BR_B}),~~M_j\ge0,~~\tr \r = 1}
\notag\\
&&
\bigg(
S
\bigg[
\sum^{m}_{j=1} (U_j\ox I_{R_B}) M_j (U_j^\dg\ox I_{R_B})
\bigg]
-S(\r)
\bigg)
\notag\\
&\ge&
K_E(U).
\eea
(ii) The first inequality in \eqref{eq:K_aep} becomes the equality if and only if there is a mixed state $\s\in \cS(\cH_{B})$, such that the equations $\tr (\s U_j^\dg U_k )=0$ hold for any $j,k$ and $j> k$.

Further, $\s$ can be chosen as diagonal if the $U_i$ are all diagonal.
$\s$ can be chosen as real if the $U_i$ are all real.
\\
(iii) $K_{Ea}(U)$ and $K_E(U)$ are both equal to $\log_2 m$ or not at the same time. If they are equal to $\log_2 m$ then the $\r$ achieving the maximum in \eqref{eq:K_aep} can be chosen as a pure state.
\\
(iv) Let $V=\sum^{m}_{j=1} Q_j \ox U_j$ be a controlled unitary on $\cH$, where the $Q_j$ are pairwise orthogonal projectors or zero projectors. Then
\bea
\label{eq:keu=kev}
&&
\log_2 m \ge \log_2 \sch (U) \ge K_E(U)\ge K_E(V),
\\
&&
\label{eq:keau=keav}
\log_2 m \ge K_{Ea}(U)\ge K_{Ea}(V).
\eea
The last equality in both equations hold when all $Q_j$ are nonzero.
\epp

The proof is given in Appendix \ref{app:{pp:aep,control}}.
If the $U_i$ in \eqref{eq:K_aep} are all diagonal then $U_i\ox I_{R_B}$ commutes with the controlled unitary $W=\sum_i \proj{i} \ox V_i$ acting on $\cH_B\ox\cH_{R_B}$ with any unitary $V_i$. The maximum in \eqref{eq:K_aep} does not change if we replace $S
\bigg[
\sum^{m}_{j=1} (U_j\ox I_{R_B}) M_j (U_j^\dg\ox I_{R_B})
\bigg]
-S(\r)$ with $S
\bigg[W
\sum^{m}_{j=1} (U_j\ox I_{R_B}) M_j (U_j^\dg\ox I_{R_B})
W^\dg\bigg]
-S(W\r W^\dg)$, Since there is no confusion, we can still name $WM_jW^\dg$ as $M_j$, and $W\r W^\dg$ as $\r$. By choosing a suitable $W$, we can assume that
the $d_{R_B}\times d_{R_B}$ diagonal blocks of any given $M_k$ are all diagonal.

The argument in \eqref{eq:K_aep} for the maximum can be replaced with $S(\r)-S
\bigg[
\sum^{m}_{j=1} (U_j^\dg\ox I_{R_B}) M_j (U_j\ox I_{R_B})
\bigg]$. It is realized by replacing $M_j$ by $(U_j^\dg\ox I_{R_B}) M_j (U_j\ox I_{R_B})$ in  \eqref{eq:K_aep}.

We have shown  in Proposition \ref{pp:aep,control} (ii) that the $\r$ by which the first inequality becomes the equality can be chosen as a pure state. For general $\r$ the proof of (ii)
implies the equations $\bra{\ps}(U_j^\dg U_k \ox I_{R_B})\ket{\ph}=0$ for any $\ket{\ps},\ket{\ph}\in\cR(\r)$. Note that
$
\rank \r=\dim\cR(\r)=\dim \bigg( (U_j^\dg U_k \ox I_{R_B})\cR(\r) \bigg).
$
If $\rank\r \ge \lf {d_Bd_{R_B}\over2}\rf + 1$, then the two subspaces $\cR(\r)$ and $(U_j^\dg U_k \ox I_{R_B})\cR(\r)$ intersect. So the equation cannot be satisfied. Hence, we have $\rank \r \le \lf {d_Bd_{R_B}\over2} \rf$.

%

The condition $\tr (\s U_j^\dg U_k )=0$ in Proposition \ref{pp:aep,control} (ii) cannot be satisfied when $\sch (U):=r< m$. To explain this fact, without loss of generality we may assume that $U_1,\cdots, U_r$ are linearly independent, and $U_{r+1}$ is the linear combination of them. Then the condition implies that $\tr \s=0$, which gives us a contradiction. So the first inequality in \eqref{eq:K_aep} is strict when $\sch (U)< m$. The inequality may be still strict when $\sch(U)=m$. An example is the $U$ whose $U_j$ are roughly equal to each other. In this case the assisted entangling power $K_{Ea}(U)$ could approach zero.

As another example, we consider the permutation unitary $U$ in Lemma \ref{le:nqp=220}.
The condition $\tr (\s U_j^\dg U_k )=0$ is equivalent to the equations
\bea
\tr (\s (I_m\op I_n \op V_1) )=0,
\\
\tr (\s (I_m\op V_3 \op I_q) )=0,
\\
\tr (\s (I_m\op V_3^\dg \op V_1) )=0.
\eea
The complex conjugate of the second equation, plus the first equation and minus the last equation results in $\tr\s=0$. This is a contradiction with the mixed state $\s$, and thus $\tr (\s U_j^\dg U_k )=0$ cannot be satisfied. So $\log_2 3 > K_{Ea}(U)\ge K_E(U)=\log_2 9 - 16/9$ ebits by Lemma \ref{le:nqp=220}. We do not know whether the inequality in $K_{Ea}(U)\ge K_E(U)$ holds for this class of $U$.

Next, it follows from Lemma \ref{le:control} that $\log_2 m$ is also the upper bound of $K_E(U)$. Proposition \ref{pp:aep,control} (iii) implies that if the bound is achievable then the entangling power and assisted entangling power are both equal to $\log_2 d_m$. It can be realized by studying the conditions in Lemma \ref{le:control}.

Finally,  Proposition \ref{pp:aep,control} (iv) does not restrict the rank of $Q_i$. It also implies that if $K_E(V)=\log_2 \sch (U)$ then the last two equalities in \eqref{eq:keu=kev} hold. For example, the $5\times2$ controlled unitary
\bea
U
&=&\proj{1} \ox I_2 + \proj{2} \ox \s_x + \proj{3} \ox \s_z
\notag\\
&+&\proj{4} \ox \bigg({\s_x+\s_z\over\sqrt2}\bigg)
+\proj{5} \ox \bigg({iI_2+\s_x\over\sqrt2}\bigg)
\eea
has Schmidt rank three. Since $V=\proj{1} \ox I_2 + \proj{2} \ox \s_x + \proj{3} \ox \s_z$ has entangling power $\log_2 3$, so does $U$. So we have provided a method of computing the entangling power of controlled unitaries whose Schmidt rank is smaller than the maximum of $d_A$ and $d_B$.
In the following two subsections, we give two families of bipartite unitaries whose assisted entangling power can be analytically derived.

\subsection{Generalized CNOT gates}
\label{subsec:gcnot}

We have described in Proposition \ref{pp:aep,control} when a controlled unitary gate has the maximum entangling and assisted entangling power. In this subsection we investigate the simplest case, namely $m=2$ in Proposition \ref{pp:aep,control}.
Let $\t_1,\cdots,\t_{d_B}$ be real numbers such that the vector $(e^{i\t_1},\cdots,e^{i\t_{d_B}})$ is orthogonal to a $d_B$-dimensional nonzero vector whose components are zeros or positive numbers.
Given a projector $P$ of rank in $[1,d_A-1]$, we say that the Schmidt-rank-two bipartite unitary gate $P\ox I_B+(I_A-P)\ox(\sum^{d_B}_{j=1}e^{i\t_j}\proj{j})$ is a \textit{generalized CNOT (GCNOT) gate} up to local unitaries. If $d_A=d_B=2$, then the definition of $\t_j$'s implies that $(e^{i\t_1},e^{i\t_2})$ is orthogonal to a two-dimensional nonzero vector whose components are zeros or positive numbers. Hence, $e^{i\t_1}=-e^{i\t_2}$. So the GCNOT gate reduces to the CNOT gate.
Using these definitions and Proposition \ref{pp:aep,control}, we can generalize Lemma 21 of \cite{cy16}.

\bpp
\label{pp:gcnot}
Let $U$ be a Schmidt-rank-two bipartite unitary. Then the following conditions are equivalent.
\\
(i) $U$ is a GCNOT gate;
\\
(ii) $K_{Ea}(U)=1$ ebit;
\\
(iii) $K_{E}(U)=1$ ebit.
\epp
\bpf
It is known that $U$ is a controlled unitary \cite{cy13}. Up to the exchange of systems and local unitaries we may assume that $U=\sum^2_{j=1} P_j \ox U_j$ as in \eqref{eq:mterms}.
The equivalence between $(ii)$ and $(iii)$ follows from Proposition \ref{pp:aep,control} (iii). Using local unitaries we may assume that $U_1=I_B$ and $U_2$ is a diagonal unitary. It does not change the entangling and assisted entangling power of $U$. If $(i)$ holds, then the definition of GCNOT gate implies that there is a diagonal density matrix
$\s$ such that $\tr (\s U_2 )=0$. So $K_E(U)=1$ ebit in terms of Proposition \ref{pp:aep,control} (ii). On the other hand,
if $(ii)$ holds, then Proposition \ref{pp:aep,control} (ii) implies that there is a diagonal density matrix
$\s$ such that $\tr (\s U_2 )=0$. So $U$ is a GCNOT gate.
We have proved $(ii)\ra(i)$.
This completes the proof.
\epf

The result shows that the GCNOT gate has the maximum entangling and assisted entangling power among Schmidt-rank-two bipartite unitaries of high dimensions. So the GCNOT gate plays the same role as the CNOT gate does in the two-qubit unitary gates. On the other hand, the GCNOT gate with dimension bigger than two contains parameters up to local unitaries, while the CNOT gate is constant. So the set of GCNOT gates contains more than one element, and this is a primary difference between the GCNOT and CNOT gates. Nevertheless, we do not know the difference between GCNOT gates in the same dimensions. On the other hand, the definition of GCNOT gates implies that if the dimension of a Schmidt-rank-two bipartite unitary is bigger, then it is more possible to become a GCNOT gate.

Proposition \ref{pp:aep,control} shows that different controlled unitaries may have the same entangling and assisted entangling power, respectively. It helps derive them for more bipartite unitaries. For Schmidt-rank-two bipartite unitaries, we can simplify their structure by the following lemma.

\bl
\label{le:coarse}
Let $U=P\ox I_{B}+(I_A-P)\ox (\sum_j e^{i\t_j} P_j)$ be a Schmidt-rank-two controlled unitary where $P$ is a projector and $P_j$ are orthogonal projectors. Let $V=\proj{1}\ox I_n + \proj{2} \ox \sum_j e^{i\t_j}\proj{j}$. Then $K_E(U)=K_E(V)$ and $K_{Ea}(U)=K_{Ea}(V)$.
\el
\bpf
Let $W=\proj{1} \ox I_{B}+\proj{2}\ox (\sum_j e^{i\t_j} P_j)$. By setting in the last statement of Proposition \ref{pp:aep,control} (iv) the $Q_j$ as rank-one projectors, we have $K_E(U)=K_E(W)$ and $K_{Ea}(U)=K_{Ea}(W)$. Up to the exchange of systems the last statement of Proposition \ref{pp:aep,control} (iv) shows that $K_E(W)=K_E(V)$ and $K_{Ea}(W)=K_{Ea}(V)$.
This completes the proof.
\epf

As a consequence of Lemma~\ref{le:coarse}, we obtain a possible simplification for the proof of Conjecture \ref{cj:sr2}: We need only consider the case that the $\t_j\in[0,2\p)$ in the conjecture are pairwise different. Finally, the generalized GCNOT gates may be defined as $\sum_j \proj{j}\ox D_j$, where each $D_j$ is a diagonal unitary such that $\tr D_j^\dg D_k=0$ for $j\ne k$. The existence of such gates is related to an open problem on the partial Hadamard matrices \cite{deLauney2010}.

\subsection{Generalized Clifford operators}
\label{subsec:clifford}

In this subsection we derive the closed formula of assisted entangling power of bipartite Clifford operators.
Let $\sigma_x,\sigma_y,\sigma_z$ be the usual $2\times 2$ Pauli matrices. Define the Pauli group $\cP_n$ to be consisting of unitary operators on $n$ qubits of the form
$e^{ik\pi/2}\bigotimes_{j=1}^n \sigma_{a_j}$,
where $a_j\in\{0,1,2,3\}$, and $\sigma_0=I_2$, $\sigma_1=\sigma_x$, $\sigma_2=\sigma_y$, and $\sigma_3=\sigma_z$, and $k$ is an integer. A unitary operator $C$ on $n$ qubits is a Clifford operator if and only if
\bea\label{eq:clifford}
C S C^\dag \in \cP_n,\quad\forall S\in\cP_n.
\eea
For example, the one-qubit Hadamard gate and the two-qubit CNOT gate are Clifford gates, but the Tofolli gate on 3 qubits is not a Clifford gate. Almost all quantum gates are not Clifford gates. The generalized Pauli group on $d$-dimensional qudits can be defined as the group generated by the following two unitary operators \cite{wm13}:
\bea\label{eq:xzdef}
X&=&\sum_{k=0}^{d-1}\ketbra{(k-1)\mbox{ mod }d}{k},\notag\\
Z&=&\sum_{k=0}^{d-1} e^{2\pi i k/d}\ketbra{k}{k}.
\eea
Then the generalized Clifford operators are defined as those $C$ which satisfy \eqref{eq:clifford} when the $\cP_n$ is understood as the generalized Pauli group on $n$ qudits.

It is claimed in \cite{wm13} that the asymptotic entanglement cost for approximately implementing two-qudit generalized Clifford gates $U$ (viewed as a bipartite unitary across the two qudits) is equal to the Schmidt strength of $U$. Better yet, the one-shot entanglement cost $E_c(U)$ for exactly implementing two-qudit generalized Clifford gates $U$ is equal to $K_{Sch}(U)$, which can be obtained by a protocol as follows (it is mentioned in Protocol 7 of \cite{yn16}, but is known before, see e.g. a more general protocol in \cite{Speelman15}): It is generalized from the protocol shown in \cite[Fig. 2]{gc99} by changing the target gate from a CNOT gate to any two-qudit Clifford gate, replacing the initial state $\ket{\chi}$ with $\frac{1}{d}\sum_{j=1}^d\sum_{k=1}^d \ket{j}_a U(\ket{j}_A \ket{k}_B)\ket{k}_b$ (the systems $a$,$A$,$B$,$b$ correspond to the four middle lines of \cite[Fig. 2]{gc99}, in the up-to-down order), changing the local Bell measurements to generalized Bell measurements, and changing the Pauli gates to generalized Pauli gates. The reason such a protocol works is that the generalized Clifford operators map the generalized Pauli operators to the generalized Pauli operators. The two qudits here are assumed to be of equal dimension, since when the dimensions are unequal, we suspect there might not be a nontrivial Clifford group. More generally, the protocol can be extended to the case that the two input systems $A$ and $B$ contain $m$ and $n$ qudits of equal dimension $d$, respectively. We call the $U$ in such general cases as a bipartite generalized Clifford operator. We have the following.

\bpp\label{prop:Clifford}
All bipartite generalized Clifford operators $V$ satisfy that
\bea\label{eq:ecbound2}
&& K_E(V)=K_{Ea}(V)=K'_{Ea}(V)=E'_c(V)=E_c(V)
\notag\\
=&&K_{Sch}(V)
=
K_d(V)
=
-\sum_{j=1}^r c_j^2 \log_2 c_j^2,
\eea
where the positive constants $c_j$ are uniquely decided by \eqref{eq:schmidtstandard}.
\epp
\bpf
The equality $E_c(U)=K_{Sch(U)}$ in the above paragraph, together with \eqref{eq:ecbound}, \eqref{eq:schmidt_strength}, and \eqref{eq:keu>=ksch} imply the assertion except $K_d(V)$. Further,
we have
\bea
\label{eq:ecbound3}
K_d(V)=K_{Ea}(V^\dg)=K_{\sch}(V^\dg)=-\sum_{j=1}^r c_j^2 \log_2 c_j^2.
\eea
The first equality follows from the definition of disentangling power. The second equality in \eqref{eq:ecbound3} follows from other equalities in \eqref{eq:ecbound2} and the fact that $V^\dg$ is also a generalized Clifford operator; the latter follows from \eqref{eq:clifford} because $C S_1 C^\dag = S_2$ is equivalent to $S_1=C^\dag S_2 C$, where $C$ is a generalized Clifford operator and $S_1,S_2$ are generalized Pauli operators. The last equality in \eqref{eq:ecbound3} follows from \eqref{eq:schmidtstandard} and \eqref{eq:schmidt_strength}.
This completes the proof.
\epf

\section{Relation between entangling and assisted entangling power}
\label{sec:eae}

We have investigated the entangling and assisted entangling power of bipartite unitaries in terms of the definitions in \eqref{eq:K_e} and \eqref{eq:K_ea}. An alternative definition of entangling power is to replace the product state in \eqref{eq:K_e} with separable states, i.e.,
$\max_{p_j,\ket{\a_j},\ket{\b_j}} E'(\sum_j p_j U\proj{\alpha_j,\beta_j}U^\dg)$, where $E'$ is a bipartite entanglement measure of systems $AR_A$ and $BR_B$. Many fundamental entanglement measures such as the entanglement of formation \cite{bds96}, the relative entropy of entanglement \cite{vp1998}, and the geometric measure of entanglement \cite{wg2003} are convex. If $E'$ is one of these measures then we have
\bea
&&\max_{p_j,\ket{\a_j},\ket{\b_j}} E'(\sum_j p_j U\proj{\alpha_j,\beta_j}U^\dg)
\notag\\
&\le&
\max_{p_j,\ket{\a_j},\ket{\b_j}} \sum_j p_j  E'(U\proj{\alpha_j,\beta_j}U^\dg)
\notag\\
&\le&
\max_{\ket{\a_j},\ket{\b_j}}  E'(U\proj{\alpha_j,\beta_j}U^\dg)
\notag\\
&=&
K_E(U).
\eea
The last equality follows from the fact that any entanglement measure reduces to the von Neumann entropy for bipartite pure states. Hence, the two definitions coincide in many cases and it suffices to use \eqref{eq:K_e} for quantifying the entangling power of bipartite unitaries.

Next we quantitatively characterize \eqref{eq:ecbound}.
\bl
\label{le:andreas}
Let $d_A\le d_B$ and $U$ a bipartite unitary. We have
\bea
\label{eq:keau>=keu}
2\log_2 d_A \ge K'_{Ea}(U) \ge K_{Ea}(U) \ge K_E(U),
\eea
and the two inequalities become equalities at the same time. When they are equalities,  the input state can be chosen as a product state $\ket{\Ps}_{AR_A}\ox\ket{\Ph}_{BR_B}$ where $\ket{\Ps}$ is the $d_A \times d_A$ maximally entangled state.
\el
\bpf
The last two inequalities in the assertion follow from \eqref{eq:ecbound}.
If the inequality $2\log_2 d_A \ge K_{Ea}(U)$ holds, then $2\log_2 d_A \ge K'_{Ea}(U)$ follows from the definition of $K'_{Ea}(U)$. The inequality holds because such number of ebits can implement $U$ by teleporting the system of Alice to Bob, performing the $U$ locally on Bob's side, and teleporting the output of system $A$ back to Alice.
This completes the proof.
\epf

Compared with the assisted entangling power, the asymptotic assisted entangling power is a tighter lower bound for the entanglement cost under LOCC.
One can show that the inequalities in \eqref{eq:keau>=keu} become equalities when $U$ is the $d_A \times d_A$ SWAP gate.
In Proposition \ref{pp:aep,control} for controlled unitaries $U$, we have shown a tighter upper bound of $K_{Ea}(U)$ than that in \eqref{eq:keau>=keu}.
So the first inequality in \eqref{eq:keau>=keu} can be strict.
On the other hand, the last inequality in \eqref{eq:keau>=keu} can also be strict by the following argument which is based on \cite[Theorem 3]{Nielsen03}. Let $U=\sqrt{1-p}I\ox I +i\sqrt{p}X\ox X$ for some $p\in [0,1]$. The proof of \cite[Theorem 3]{Nielsen03} shows that $K_E(U\ox U)\ge H[(1-2p)^2]$, as well as the fact that $H[(1-2p)^2]>2H(p)$ for some range of $p$. It is shown in the proof of \cite[Theorem 2]{Nielsen03} that $K_E(U)=H(p)$. Hence, for $p$ in some range, the strict inequality $K_E(U\ox U)>2K_E(U)$ holds, and by definition $K_{Ea}(U)\ge\frac{1}{2}K_E(U\ox U)$; thus, for some two-qubit Schmidt-rank-two unitaries $U$, the strict inequality
$
K_{Ea}(U)>K_E(U)
$
holds. Note that \cite{Nielsen03} does not mention this inequality (although the above argument means that this inequality is essentially implied by their analysis), but it remarks that the inequality $K_{\Delta E}(U)>K_E(U)$ holds for some $U$; see the comment on \cite[p5]{Nielsen03}, which is based on \cite[Theorem 3]{Nielsen03}. The latter inequality is a weaker inequality because $K_{\Delta E}(U)\ge K_{Ea}(U)$ for any bipartite unitary $U$.

Next we investigate the disentangling power using Lemma \ref{le:andreas}.


\subsection{Disentangling power}
\label{subsec:dis}

The example with different disentangling power and assisted entangling power in \cite{lsw09} is a Schmidt-rank-four $3\times2$ bipartite unitary. We show that many bipartite unitaries of smaller Schmidt rank have equal disentangling power and assisted entangling power. Note that the complex conjugate of a complex permutation unitary is still a complex permutation unitary. From Proposition \ref{pp:dax2} and Lemma \ref{le:andreas} we get the following.
\bt
\label{thm:dax2}
Any $2\times d_B$ complex permutation unitary $U$ of Schmidt rank four satisfies
\bea
K_d(U) = K_{Ea}(U)= K_E(U) = 2
\eea
ebits.
\et

To construct more examples, we present a preliminary lemma.
The definitions of $K_{E}(U)$ and $K_{Ea}(U)$ imply
\bea
\label{eq:ke(u)}
K_{E}(U)=K_{E}(U^*),
\notag\\
K_{Ea}(U)=K_{Ea}(U^*),
\eea
and thus we are led to the following lemma.
\bl
\label{le:u=ut}
$K_E(U)=K_E(U^\dg)$ and $K_{Ea}(U)=K_{Ea}(U^\dg)$ holds when the bipartite unitary $U$ is locally equivalent to $U^\dg$ or a symmetric matrix.
\el
For example, such $U$ can be any two-qubit unitary, because it is the sum of the tensor product of Pauli operators. Next, $U$ can also be
any Schmidt-rank-two bipartite unitary because it is locally equivalent to the diagonal unitary. A nontrivial example is as follows.
\bpp
\label{pp:dx2,sr3}
Any Schmidt-rank-three $d_A\times 2$ bipartite unitary is locally equivalent to a symmetric matrix.
\epp
\bpf
Let $U$ be a  Schmidt-rank-three $d_A\times 2$ unitary.
It is known that $U$ is a controlled unitary \cite{cy14}. Up to local unitaries we may assume that $U=\sum^{d_A}_{j=1} \proj{j}\ox U_j$ where $U_j$ are all $2\times2$ unitary matrices, and the first three of them are linearly independent.
Up to local unitaries on $\cH_B$ we may assume that $U_1=I_2$ and $U_2$ is diagonal. Since $U_3$ is a $2\times2$ unitary,
the non-diagonal entries of $U_3$ have the same
modulus. We can perform suitable diagonal local unitaries to make the two
entries equal to the modulus. Then the resulting $U_1, U_2$ and $U_3$ are all
symmetric. Since any $U_j$ is the linear combination of them, it is also symmetric. Hence, $U$ is locally equivalent to a symmetric matrix. This completes the proof.
\epf

Equation \eqref{eq:ke(u)} implies that performing the complex conjugate on any bipartite unitary operation does not change the entangling power, assisted entangling power, and disentangling power of this operation. This phenomenon
could still hold for multipartite unitaries if we generalize the definition of the three powers to multipartite scenario. So we may regard the complex conjugate as a local operation for nonlocal unitaries $U$, though the $U^*$ is generally not convertible to $U$ via LOCC. Nevertheless, they are convertible via stochastic LOCC. Since $U$ of Schmidt rank $r$ can be used to generate a Schmidt-rank-$r$ entangled state,
which can be converted with some probability into a uniformly entangled state of Schmidt rank $r$ implementing $U^\ast$ probabilistically. We construct the protocol for the implementation in the next subsection.

\subsection{A probabilistic protocol for implementing bipartite unitaries}

Given a bipartite unitary $U$, we may assume $U=\sum_{j=1}^{r} c_j A_j\ox B_j$, where $r$ is the Schmidt rank of $U$, and
$\tr(A_j^\dag A_k)=\delta_{jk}=\tr(B_j^\dag B_k)$, where $\delta$ is the Kronecker delta symbol, and $c_j$ are positive coefficients. The unitarity condition $U^\dag U=I_{AB}$ implies that
\bea
I_{AB}=\sum_{j,k=1}^r c_k^\ast c_j A_k^\dag A_j \ox B_k^\dag B_j
\eea
Taking partial trace over system $A$, we have
$
I_B=\frac{1}{d_A}\sum_{j=1}^r c_j^2 B_j^\dag B_j
$
Similarly,
$
I_A=\frac{1}{d_B}\sum_{j=1}^r c_j^2  A_j^\dag A_j$.
Hence $\{\frac{c_j}{\sqrt{d_A}} A_j\}$ could be a set of Kraus operators for a quantum channel on system $A$,
and $\{\frac{c_j}{\sqrt{d_B}} B_j\}$ could be a set of Kraus operators for a quantum channel on system $B$.

The unitary $U$ can be implemented with some probability using the following protocol. Assume the input state is $\ket{\Psi}_{AB}$.
Suppose there is a Schmidt-rank-$r$ entangled resource state $\ket{\psi}=\frac{1}{\sqrt{r}}\sum_{j=1}^r \ket{j}_e\ket{j}_f$, where $e$ and $f$ are $r$-dimensional ancillary systems on the $A$ and $B$ side, respectively. The protocol also uses $r$-dimensional ancillary systems $a$ and $b$ on the $A$ and $B$ side, respectively. The $a$ and $b$ are initialized in the state $\ket{0}$.

1. Perform a local unitary on $Aa$ that implements a quantum channel on $A$ with $\frac{c_j}{\sqrt{d_A}} A_j$ as Kraus operators, so that the output of $a$ in its computational basis contains full information about which Kraus operator was applied on $A$. (However, we do not perform a measurement on $a$ at this stage.) Similarly, perform a local unitary on $Bb$, which implements a quantum channel on $B$ with $\frac{c_j}{\sqrt{d_B}} B_j$ as Kraus operators.

2. Perform a local controlled-cyclic-shift gate on $ae$, and measure $e$ in the computational basis. Similarly, perform a local controlled-cyclic-shift gate on $bf$, and measure $f$ in the computational basis. Perform a Fourier gate on $a$ and then measure $a$ in the computational basis. Perform a different unitary on $b$ with the first row in its matrix proportional to $(1/c_1,\dots,1/c_r)$, and then measure $b$ in the computational basis.

In general, the protocol implements the nonlocal unitary $U$ with probability $1/r^3$. However, when $c_j$ are all equal, the above procedure implements the (possibly non-unitary) operator $V_{lm}=\sum_{j=1}^r e^{2\pi i m j/r} A_j \ox B_{1+(j+l-2)\mod r}$ for $l,m\in\{1,\dots,r\}$, and $l,m$ take their possible values with equal probabilities; hence, the success probability is $1/r^2$ in this case.

\section{Two conjectures}
\label{sec:open}

In this section we discuss two conjectures, respectively arising in the literature and this paper. The first conjecture is related to the dimension of reference systems of input states saturating the assisted entangling power. The second conjecture is to construct the upper bound of assisted entangling power in terms of the Schmidt rank of input bipartite unitaries. They both aim for a further understanding of the assisted entangling power.

\subsection{The dimension of reference systems of assisted entangling power}
\label{subsec:ancillasize}

If $\ket{\ps}_{AR_A:BR_B}$ maximizes the function $E(U(\ket{\psi}))-E(\ket{\psi})$ of \eqref{eq:K_ea}, then we call it the assisted state of $U$, and the dimensions of $R_A$ and $R_B$ are respectively denoted as $d_{R_A}$ and $d_{R_B}$. Let $R_A''=R_AR_A'$ and $R_B''=R_BR_B'$ be a bigger reference system and $\ket{\ph}$ a pure state of the system $R_A'R_B'$. Then the state $\ket{\ps}\ox\ket{\ph}\in\cH_{R_A''R_B''}$ is another assisted state of $U$. Hence, there are infinitely many assisted states, and the dimension of reference system can be arbitrarily large. On the other hand, it is an open problem to derive the minimum dimension of reference system, respectively denoted as $d^{opt}_{R_A}$ and $d^{opt}_{R_B}$.  As the remarks in \cite{Nielsen03} suggest, it is an open problem to find $d^{opt}_{R_A}$ (or $d^{opt}_{R_B}$) as a function of $d_A, d_B$ only, or as a function of $U$ for generic $U$ with a fixed pair of $(d_A,d_B)$. So far there is no evidence whether $d^{opt}_{R_A}$ is finite. To estimate the assisted entangling power,
it was asked whether \cite{privateandreas}
\bea
\label{eq:ra<=da}
d^{opt}_{R_A} \le d_A,
\\
\label{eq:rb<=db}
d^{opt}_{R_B} \le d_B.
\eea
If $U$ is a controlled unitary controlled from the $A$ side, then
$d^{opt}_{R_B}$ could be at most $d_B$ by Proposition \ref{pp:aep,control} (ii). It is a hint to the above conjecture.
Since $d_A,d_B,d^{opt}_{R_A}$, and $d^{opt}_{R_B}$ are from the same pure state $\ket{\ps}$, we have
$\sr(\ps)\le \min\{d_Ad^{opt}_{R_A},d_Bd^{opt}_{R_B}\}$, where $\sr$ means the Schmidt rank.
We do not know whether the two conjectured inequalities \eqref{eq:ra<=da} and \eqref{eq:rb<=db} are independent.

\subsection{The upper bound of assisted entangling power} \label{subsec:keau<=log2schu}

In Proposition \ref{pp:aep,control}, we have obtained an upper bound of assisted entangling power of bipartite controlled unitaries. We present a conjecture similar to Lemma \ref{le:andreas}.
\bcj
\label{cj:andreas}
Let $U$ be a bipartite unitary. We have
\bea
\label{eq:keaulelogr}
\log_2 \sch (U) \ge K_{Ea}(U)\ge K_E(U),
\eea
and the two inequalities become equalities at the same time. When they are equalities,  the input state can be chosen as a product state $\ket{\Ps}_{AR_A}\ox\ket{\Ph}_{BR_B}$.
\ecj

It would be a tighter upper bound than the first inequality in \eqref{eq:keau>=keu}, because $\sch (U) \le \min \{d_A^2,d_B^2\} $.
Note that if the assisted entangling power is replaced with the entangling power, then the conjecture holds by definition.
We provide a few evidences supporting the conjecture.
The inequality holds for any two-qubit unitary $U$, whose Schmidt rank can be 1,2, or 4 \cite{Nielsen03}. If $\sch(U)=2$, then the inequality follows from Lemma \ref{le:keu-sr2complex}. If $\sch(U)=4$, then the inequality follows from Lemma \ref{le:andreas}.
If $U=\sum_j P_j\ox U_j$ is controlled with $m$ terms, then Proposition \ref{pp:aep,control} (i) implies that the inequality in \eqref{eq:keaulelogr} holds when the $U_j$ are linearly independent.
We prove a special case of \eqref{eq:keaulelogr}.
Suppose the assisted state of $U$ on $\cH_{AB}$ can be written as
\bea
\label{eq:ketps=sqrta}
\ket{\ps}=\sqrt{a}\ket{\m}_{AR_A:BR_B}+\sqrt{1-a}\ket{\n}_{AR_A:BR_B},
\eea
where $a\in[0,1]$, and $\ket{\m}$, $\ket{\n}$ are orthogonal product states,  and the $R_B$ space of $\ket{\m}$ is orthogonal to that of $\ket{\n}$. Then
\bea
\label{eq:kea(u)}
&&
K_{Ea}(U)
\notag\\
&=&
E(U\ket{\psi})-E(\ket{\psi})
\notag\\
&=&
S(\tr_{BR_B} U\proj{\psi}U^\dg )
-
S(\tr_{BR_B} \proj{\psi} )
\notag\\
&=&
S(a \tr_{BR_B} U\proj{\m}U^\dg + (1-a) \tr_{BR_B} U\proj{\n}U^\dg )
\notag\\
&&-
S(\tr_{BR_B} \proj{\psi} )
\notag\\
&\le&
a S(\tr_{BR_B}U\proj{\m}U^\dg) + (1-a) S (\tr_{BR_B}U\proj{\n}U^\dg)
\notag\\
&&+
H(a,1-a)
-
S(\tr_{BR_B} \proj{\psi} )
\notag\\
&\le&
a S(\tr_{BR_B}U\proj{\m}U^\dg) + (1-a) S (\tr_{BR_B}U\proj{\n}U^\dg)
\notag\\
&\le&
\log_2 \sch (U).
\eea
The first inequality follows from Lemma \ref{le:concave} (ii). The second inequality follows from Lemma \ref{le:majorization} because the vector $\des(a,1-a)$ is majorized by the Schmidt vector of $\ket{\ps}$ by \cite[Corollary 4]{Nielsen00}. The last inequality follows from the fact that $\ket{\m}$, $\ket{\n}$ are both product states.

If the $R_B$ space of $\ket{\m}$ is not orthogonal to that of $\ket{\n}$, then we construct another pure state
\bea
\label{eq:ketps=sqrta-}
\ket{\ph}=\sqrt{a}\ket{\m}_{AR_A:BR_B}-\sqrt{1-a}\ket{\n}_{AR_A:BR_B}.
\eea
Then
\bea
&&
\min_{x=U\ket{\ps},U\ket{\ph}} E(x)
\notag\\
&\le&
{1\over2} E (U\ket{\ps}) + {1\over2} E (U\ket{\ph})
\notag\\
&\le&
S
(
{1\over2} \tr_{BR_B} U\proj{\ps}U^\dg
+
{1\over2} \tr_{BR_B} U\proj{\ph}U^\dg
)
\notag\\
&=&
S
(
a \tr_{BR_B} U\proj{\m}U^\dg
+
(1-a) \tr_{BR_B} U\proj{\n}U^\dg
)
\notag\\
&\le&
a S( \tr_{BR_B} U\proj{\ps}U^\dg)
+
(1-a) S ( \tr_{BR_B} U\proj{\ph}U^\dg)
\notag\\
&&+
H(a,1-a)
\notag\\
&\le&
\log_2 \sch (U)
+
E(\ps).
\eea
The inequalities hold by arguments similar to that for \eqref{eq:kea(u)}. Since $E(\ps)=E(\ph)$, we have
\bea
&&
\min\{E(U\ket{\ps})-E(\ket{\ps}),
E(U\ket{\ph})-E(\ket{\ph})\}
\notag\\
&\le&
\log_2 \sch(U).
\eea
However we do not know whether the inequality holds when the minimum is replaced with the maximum.

\section{Conclusions}
\label{sec:con}

In this paper we have analytically derived the entangling power of Schmidt-rank-two bipartite unitary, Schmidt-rank-three permutation unitary and some special non-controlled unitary operations. In particular the entangling power of any bipartite permutation unitary of Schmidt rank three can only take one of two values: $\log_2 9 - 16/9$ or $\log_2 3$ ebits. We have proposed the upper bound of the assisted entangling power of bipartite controlled unitaries, and the necessary and sufficient conditions for this upper bound.
The entangling power, assisted entangling power and disentangling power of $2\times d_B$ permutation unitaries of Schmidt rank four are all $2$ ebits. These quantities are also derived for generalized Clifford operators.
We further show that any bipartite permutation unitary of Schmidt rank greater than two has entangling power greater than $1.223$ ebits.

We also have constructed GCNOT gates, which is a parameterized Schmidt-rank-two bipartite unitary whose assisted entangling power is 1 ebit. It generalizes the known CNOT gate for two-qubit systems.
Further we have constructed the inequalities between entangling power and assisted entangling power, and conditions by which the inequalities hold.
We also have shown the connection to the disentangling power by proposing a probabilistic protocol for implementing bipartite unitaries. The next step is to analyze the conjectures in Sec.~\ref{sec:open}. By studying the properties of the different types of entangling power, we hope to get more insight into the question of whether there is a bipartite unitary such that its entanglement cost is strictly greater than its assisted entangling power.

\section*{Acknowledgments}

L.C. was supported by the NSF of China (Grant No. 11501024), and the Fundamental Research Funds for the Central Universities (Grants No. 30426401 and No. 30458601). L.Y. was supported by NICT-A (Japan).

\appendix

\section{The proof of Lemma~\ref{le:control}}
\label{app:{le:control}}

\bpf
We prove the assertions when all $P_j$ have rank one. One can similarly prove the assertions.
\\
(i)
Let $\ket{\a,\b}$ be the critical state of $U$, where $\ket{\a}=\sum_j \sqrt{p_j}\ket{j,a_j}_{AR_A}$,  $\sum_j p_j=1$, and $p_j>0$. Hence,
\bea
\label{eq:keu(i)}
K_E(U)
&=& E (\sum_j \sqrt{p_j}\ket{j,a_j}_{AR_A} \ox (U_j)_B \ket{\b}_{BR_B})
\notag\\
&=& E (\sum_j \sqrt{p_j}\ket{j}_{A} \ox (U_j)_B \ket{\b}_{BR_B})
\notag\\
&=& E (U \sum_j \sqrt{p_j}\ket{j}_{A} \ox \ket{\b}_{BR_B})
\notag\\
&\le& \max_{\ket{\alpha}\in\cH_A,\ket{\beta}\in\cH_{BR_B}} E(U(\ket{\alpha}\ket{\beta}))
\notag\\
&\le& K_E(U).
\eea
The second equality follows from the fact that local unitaries does not change the amount of entanglement.
Hence, the first equality in \eqref{eq:K_eu} follows. In particular, $\sum_j \sqrt{p_j}\ket{j}_{A} \ox \ket{\b}_{BR_B}$ is another critical state of $U$.

The second equality in \eqref{eq:K_eu} follows from the definition of $E$ and the assumption $\ket{\a}= \sum^{d_A}_{j=1} \sqrt{p_j}\ket{j}$, where $p_j\ge0,~\sum^{d_A}_{j=1} p_j=1$.
To prove the inequalities in \eqref{eq:K_eu}, we note that $K_E(U)\le\log_2 \text{Sch}(U)$ is the definition of $K_E$. The last inequality of \eqref{eq:K_eu} follows from the definition of $U$. The last assertion of (i) holds because $U(\ket{\a,\b})$ has Schmidt rank at most $\sch(U)$.

(ii) The proof is similar to that of (i). In particular, the first equality in \eqref{eq:K_euab} follows by applying \eqref{eq:K_eu} to both systems of $U$.

(iii) Equation \eqref{eq:K_euab2} is trivial. An example for which the inequality holds is $U=\sum_{j=0}^3 \ketbra{j}{j} \ox \sigma_j$. The entangling power with a one-qubit ancilla $R_B$ initially maximally entangled with $B$ is 2 ebits, while the entangling power without $R_B$ is 1 ebit.

On the other hand, an example for which the inequality does not hold is $U=\sum_{j=1}^{d_A} \proj{j}\ox V_j$, where $d_A\le d_B$, $V_j$ are $d_B\times d_B$ permutation matrices whose $(j,1)$ element is $1$, and at the same time they do not have simultaneous singular value decomposition. One can easily verify that such $V_j$ exist. So $U$ is not locally equivalent to a controlled unitary from the $B$ side. Evidently, $U$ has Schmidt rank $d_A$, and thus $K_E(U)\le \log_2 d_A$. This upper bound is achievable, and a critical state of $U$ is the input state $(\frac{1}{\sqrt{d_A}}\sum_{j=1}^{d_A} \ket{j}_A) \ox \ket{1}_B$. Using these results, \eqref{eq:K_eu} and \eqref{eq:K_euab2} we have
\bea
\log_2 d_A
&\ge&
K_E(U)
\notag\\
&=&
\max_{\ket{\alpha}\in\cH_A,\ket{\beta}\in\cH_{BR_B}} E(U(\ket{\alpha}\ket{\beta}))
\notag\\
&\ge&
\max_{\ket{\alpha}\in\cH_A,\ket{\beta}\in\cH_{B}} E(U(\ket{\alpha}\ket{\beta}))
\notag\\
&=&
\log_2 d_A.
\eea
Hence the equality in \eqref{eq:K_euab2}  holds.

(iv) The assertion follows from \eqref{eq:keu(i)} and \eqref{eq:K_euab}.
This completes the proof.
\epf

\section{The proof of Lemma~\ref{le:nqp=220}}
\label{app:{le:nqp=220}}

\bpf
Since $U$ is unitary, $V_1$ and $V_3$ are also unitary matrices. So $U$ is a controlled unitary controlled from both $A$ and $B$ sides. Applying Lemma \ref{le:control} to \eqref{eq:perm_u_3terms}, we have
\bea
\label{eq:nqp=220}
K_E(U)
&=&\max_{p_j\ge0,~\sum^{3}_{j=1} p_j=1,~\ket{\beta}\in\cH_{B}}
\notag\\
&&S
(p_1\proj{\b}  + p_2 \proj{\b_1} + p_3 \proj{\b_2}),
\eea
where $\ket{\b_1}=(I_m \op I_n \op V_1)\ket{\b}$ and $\ket{\b_2}=(I_m \op V_3 \op I_p)\ket{\b}$. There is a  unitary $W=W_1\op W_2 \op W_3$, such that
\bea
&&
W\ket{\b}=(a,0,\cdots,0,b,0,0,\cdots,0,c,0,0,\cdots,0),
\notag\\
&&
W\ket{\b_1}=(a,0,\cdots,0,b,0,0,\cdots,0,c_1,c_2,0,\cdots,0),
\notag\\
&&
W\ket{\b_2}=(a,0,\cdots,0,b_1,b_2,0,\cdots,0,c,0,0,\cdots,0),
\notag\\
\eea
where $\abs{b_1}^2+\abs{b_2}^2=\abs{b}^2$ and $\abs{c_1}^2+\abs{c_2}^2=\abs{c}^2$. Let $X=I_m\op X_1 \op I_{n-2} \op X_2 \op I_{q-2}$ be a unitary operator with $2\times2$ unitary matrices $X_1,X_2$. We can find an $X$ such that the first entry of $X_1(b_1,b_2)^T$ is the same as that of $X_1(b,0)^T$, and the first entry of $X_2(c_1,c_2)^T$ is the same as that of $X_2(c,0)^T$. Now we can find a suitable unitary $Y$ such that each of the three states $YXW\ket{\b}$, $YXW\ket{\b}$, and $YXW\ket{\b}$ contains exactly three nonzero entries. They are in the same rows of the three states.
Let $\ket{\b'}$, $\ket{\b_1'}$ and $\ket{\b_2'}$ be qutrits which respectively consist of the three nonzero entries. We may assume
$
\ket{\b'}=(a,b',c')^T,
\ket{\b_1'}=(a,b',c'e^{i\a})^T,
$ and $
\ket{\b_2'}=(a,b'e^{i\b},c')^T.
$
Since the von Neumann entropy is invariant up to unitary transformation, \eqref{eq:nqp=220} implies that
\bea
\label{eq:nqp=220-1}
K_E(U)
&=&\max_{p_j\ge0,~\sum^{3}_{j=1} p_j=1,~\abs{a}^2+\abs{b'}^2+\abs{c'}^2=1,~c\ge0}
\notag\\
&&S
(\r),
\eea
where the state $\r=p_1\proj{\b'}  + p_2 \proj{\b_1'} + p_3 \proj{\b_2'}$. By computation we have $\det \r = p_1p_2p_3 \abs{ab'c'(1-e^{i\a})(1-e^{i\b})}^2 $. It follows from the restriction in \eqref{eq:nqp=220-1} that $\det \r \le {16\over 729}$. Hence,
\bea
\label{eq:nqp=220-2}
K_E(U) \le \max_{\s\ge0,~\tr\s=1,~\rank\s\le3,~\det\s\le {16\over 729}} S(\s).
\eea
Let the three eigenvalues of $\s$ be $\l_1,\l_2$ and $\l_3$  in the ascending order. Since $\det\s\le {16\over 729}$, the eigenvalues cannot be all equal.
That is, we have either $\l_1<\l_2$ or $\l_2<\l_3$.

Assume that the maximum of \eqref{eq:nqp=220-2} is achieved when $\det\s< {16\over 729}$. If $\l_1<\l_2$, then we can find a small $\e>0$ to construct a quantum state $\s'$ of three eigenvalues $\l_1+\e$, $\l_2-\e$, and $\l_3$ still in the ascending order, and at the same time $\det \s' \le {16\over 729}$. Since $\s'\prec_s\s$, it follows from Lemma \ref{le:majorization} that $S(\s')>S(\s)$. It gives us a contradiction with the assumption. One may similarly find the contradiction when $\l_2<\l_3$. So the maximum of \eqref{eq:nqp=220-2} is achievable when $\det\s= {16\over 729}$. Since $\det\s=\l_1\l_2\l_3$ and $\sum^3_{j=1} \l_j =1$, using the inequality $\l_2+\l_3\ge2\sqrt{\l_2\l_3}$ we obtain
\bea
1-\l_1\ge{8\over27\sqrt{\l_1}}.
\eea
The inequality holds only if $\l_1\ge1/9$. Using the upper bound $\l_1\le1/3$,
one can plot $S(\s)$ as the function of $\l_1$ and show that the maximum is achievable when $\l_1=1/9$, $\l_2=\l_3=4/9$. It follows from
\eqref{eq:nqp=220-2} that $K_E(U)\le \log_2 9 - 16/9$. The equality holds for the permutation unitaries that fit into the form \eqref{eq:perm_u_3terms}, as shown in the proof of \cite[Proposition 1]{cy16}.
This completes the proof.
\epf

\section{The proof of Proposition~\ref{pp:cp3}}
\label{app:{pp:cp3}}

\bpf
Since $U$ has Schmidt rank three we have $n\in[1,d_B-2]$. Suppose the input state is $\ket{\a,\b}_{AR_A,BR_B}=\sum^2_{j=1}a_j\ket{j,\a_j}_{AR_A}\ox\sum^{d_B}_{k=1}b_k\ket{k,\b_k}_{BR_B}$. The entangling power of $U$ is equal to the maximum amount of entanglement contained in the state
\bea
\label{eq:uabara}
&&
U\ket{\a,\b}_{AR_A,BR_B}
\notag\\
&=&
\sqrt{x}\sum^2_{j=1}a_j\ket{j,\a_j}_{AR_A}
\sum^n_{k=1}{c_k}\ket{k,\b_k}_{BR_B}
\notag\\
&+&
a_2\sqrt{1-x}\ket{1,\a_2}_{AR_A}
\sum^{d_B}_{k=n+1}{c_k}\ket{k,\b_k}_{BR_B}
\notag\\
&+&
a_1\sqrt{1-x}\ket{2,\a_1}_{AR_A}
\sum^{d_B}_{k=n+1}{c_k}C\ket{k}\ket{\b_k}_{BR_B}
\notag\\
&:=&
\sqrt{x} \ket{\ps_1} + a_2\sqrt{1-x} \ket{\ps_2} + a_1 \sqrt{1-x}\ket{\ps_3},
\eea
where the parameters $x=\sum^n_{k=1}\abs{b_k}^2$, and $c_k={b_k\over \sqrt{x}}$ for $k\le n$, and $c_k={b_k\over \sqrt{1-x}}$ for $k>n$. Besides, the three product states $\ket{\ps_1},\ket{\ps_2}$, and $\ket{\ps_3}$ are pairwise orthogonal. We have
\bea
\label{eq:keu=s}
&&K_E(U)
=
\max_{x,a_j,\a_j,c_k,\b_k}
S
\bigg(
x\tr_{BR_B} \proj{\ps_1} +
\notag\\
&&
(1-x) \tr_{BR_B}
(a_2 \ket{\ps_2} + a_1 \ket{\ps_3})
(a_2^* \bra{\ps_2} + a_1^* \bra{\ps_3})
\bigg)
\notag\\
&&\le
\max_{x,a_j,\a_j,c_k,\b_k}
[
H(x,1-x)+
\notag\\
&&
(1-x)
S
\bigg( \tr_{BR_B}
(a_2 \ket{\ps_2} + a_1 \ket{\ps_3})
(a_2^* \bra{\ps_2} + a_1^* \bra{\ps_3})
\bigg)]
\notag\\
&&=
\max_{x,a_j,c_k,\b_k}
[H(x,1-x)+
\notag\\
&&
(1-x)
S
\bigg( \tr_{AR_A}
(\abs{a_2}^2 \proj{\ps_2} + \abs{a_1}^2\proj{\ps_3})
\bigg)].
\notag\\
\eea
The inequality follows from Lemma \ref{le:concave}, and the last equality follows from \ref{eq:uabara}. The inequality becomes the equality when $\ket{\a_1}$ and $\ket{\a_2}$ are orthogonal. This is achievable, because $\ket{\a_1}$ and $\ket{\a_2}$ do not appear in the von Neumann entropy of the final equation of \eqref{eq:keu=s}. The entropy is upper bounded by the entangling power of $V=\ketbra{1}{2}\ox I_B + \ketbra{2}{1} \ox C$, which can be obtained
using the paragraph above Proposition \ref{pp:db=3}.
Let the entangling power of $V$ be a positive constant $M\le1$. Thus, $K_E(U) \le H(x,1-x)+(1-x)M$. It is maximized at $x={1\over e^M+1}$ by considering the behavior of its first derivative in the whole range $(0,1)$. We have thus obtained the assertion. This completes the proof.
\epf

\section{The proof of Proposition~\ref{pp:>1ebit}}
\label{app:{pp:>1ebit}}

\bpf
Assume that the claim holds for any (and all) bipartite permutation unitary which is not BCPU from the $A$ side. (The definition of BCPU is just above Lemma~\ref{le:bcu}.) We assert that under such assumption, the claim holds for any $U$ which is a BCPU from the $A$ side, say $U=(\op_j)_A V_j$, where the $d_j \times d_B$ bipartite unitary $V_j$ is not a BCPU from the $A$ side, and $\sum_j d_j=d_A$. If one of the $V_j$'s has Schmidt rank greater than two, then the assertion follows from Lemma \ref{le:bcu} and the assumption. So any $V_j$ has Schmidt rank of at most two, and it is a controlled unitary \cite{cy13}. If the A-direct sum of some $V_j$'s has Schmidt rank three then the assertion follows from Lemma \ref{le:bcu} and Proposition \ref{pp:entpower3}. So it suffices to consider the case that $k$ terms of $V_j$'s ($k\ge 2$) each have Schmidt rank one or two and their A-direct sum has Schmidt rank four. Suppose $k$ is the minimum integer such that the previous sentence holds. Suppose $k\ge 3$; then under the condition established above that the A-direct sum of any set of $V_j$'s has Schmidt rank not equal to three, it must be that any $k-1$ terms in the $k$ terms satisfy that their $A$-direct sum has Schmidt rank two, and we may view these terms as one $V_j$ in the argument below. Then it suffices to prove the assertion when $k=2$, i.e., when $U=V_1\op_A V_2$ has Schmidt rank four, where $V_1$ and $V_2$ are of Schmidt rank two. From \cite[Lemma 15(i)]{cy16}, any bipartite permutation unitary of Schmidt rank two is equivalent under local permutation unitaries to a controlled-permutation unitary with two terms, where the direction of control may be from either side. If one of $V_1$ and $V_2$ is controlled from the $A$ side with two terms, then the assertion follows again from Lemma \ref{le:bcu} and Proposition \ref{pp:entpower3}.

So we may assume $V_j=W_j\ox P_j + X_j \ox Q_j$ for $j=1,2$, where $W_j$ and $X_j$ are the direct sum of a permutation matrix of order $d_j$ with a zero matrix of order $d_{3-j}$, and $P_j$ and $Q_j$ are two partial permutation matrices of order $d_B$ such that $P_j+Q_j$ is a permutation matrix. From $U=V_1\op_A V_2$, there is no common nonzero row or column for the pair of matrices $W_1$ and $W_2$, and the same holds for the pairs of matrices $(W_1, X_2)$, $(X_1,W_2)$, and $(X_1,X_2)$. Since $U=V_1\op_A V_2$ has Schmidt rank four, the four matrices $P_1,P_2,Q_1,Q_2$ are linearly independent. We can find a Schmidt-rank-two uniformly entangled state $\ket{\a}_{AR_A}$ such that the four states $(Y_j\ox I_{R_A})\ket{\a}_{AR_A}$ are pairwise orthogonal, where $Y_j=W_1,X_1,W_2$, and $X_2$; a type of choice of such state is given by $\frac{1}{\sqrt{2}}(\ket{j}_A \ox \ket{1}_{R_A} + \ket{k}_A \ox \ket{2}_{R_A})$, where $\ket{j}_A$ and $\ket{k}_A$ are computational basis states, and $W_1\ket{j}_A$ and $X_1 \ket{j}_A$ are nonzero and orthogonal to each other, and $W_2\ket{k}_A$ and $X_2 \ket{k}_A$ are nonzero and orthogonal to each other, and $W_2\ket{j}_A=X_2\ket{j}_A=W_1\ket{k}_A=X_1\ket{k}_A=0$. We can also find another Schmidt-rank-two uniformly entangled state $\ket{\b}_{BR_B}$ such that three of the four states $(Z_j\ox I_{R_B})\ket{\b}_{BR_B}$ are pairwise orthogonal where $Z_j=P_1,Q_1,P_2$, and $Q_2$. The fourth state is either the same as one of the three states, or orthogonal to all of them. Let $\ket{\a}_{AR_A}\ox\ket{\b}_{BR_B}$ be the input state; then the corresponding amount of output entanglement is either $2$ ebits or $-{2\over3}\log_2{2\over3}-2\times{1\over6}\log_2 {1\over6}=\log_2 3-\frac{1}{3}>1.251$ ebits. So the assertion holds. From now on we assume that $U$ is not a BCPU. If $U$ is of Schmidt rank three, the claim already follows from Proposition~\ref{pp:entpower3}. Thus, in the following we assume $U$ is of Schmidt rank at least four.


Suppose $U$ contains at least three nonzero blocks in one big column. We may assume that the big column is the first big column in $U$, and its first three blocks are nonzero. Up to local permutations on $B$ we may assume that the $j$'th column of the $j$'th block is nonzero for $j=1,2,3$. Let the initial state on $A R_A$ be a product state $\ket{1}_A \ket{1}_{R_A}$, and let the initial state on $B R_B$ be $\ket{\phi}_{B R_B}=\frac{1}{\sqrt{3}}\sum_{j=1}^3 \ket{j}_B \ket{j}_{R_B}$. We obtain that the output entanglement is exactly $\log_2 3$ ebits and the assertion holds.

Hence, $U=\sum^{d_A}_{j,k=1}\ketbra{j}{k}\ox U_{jk}$ has exactly two nonzero blocks in every big column. We consider the four blocks that are the intersections of two big rows and two big columns in $U$. Denote $V$ as the submatrix formed by these four blocks. We have $V=\sum^2_{j,k=1} \ketbra{j}{k}\ox U_{jk}$, where each $U_{jk}$ is a $d_B\times d_B$ partial permutation matrix.

Suppose there is a $V$ such that all four blocks in it are nonzero. Since each big column of $V$ contains the only nonzero blocks in the corresponding big column of $U$, all columns of $V$ are nonzero. Thus, $V$ contains $2d_B$ nonzero elements; hence, all rows of $V$ are nonzero, and $V$ is a permutation matrix. Since $U$ is not a BCPU, we obtain $U=V$. Then since we assumed previously that $U$ is of Schmidt rank at least four, the Schmidt rank of $U$ is exactly four. The assertion follows from Proposition \ref{pp:dax2}.

%
%

It remains to consider the case that any $V$ contains at most three nonzero blocks. The above assumptions imply that there is a $V$ containing exactly three nonzero blocks. Assume that any $V$ has Schmidt rank smaller than three. Then up to local permutation matrices we may assume that $U_{11}=I_r \op 0_{d_B-r}$ and $U_{12}=U_{21}=0_r \op P$, where $r\in[1,d_B-1]$ and $P$ is a permutation matrix. Up to local permutation unitaries we may assume $U_{23}\ne 0$ and $U_{32}\ne 0$.
Applying the above assumption about $V$ to the four blocks $U_{11},U_{12},U_{31},U_{32}$, we get that $U_{32}=U_{11}$. Similarly, $U_{23}=U_{11}$. Up to local permutation unitaries we may assume $U_{34}\ne 0$ and $U_{43}\ne 0$.  Applying the above assumption about $V$ to the four blocks $U_{12},U_{14},U_{32},U_{34}$, we have $U_{34}=U_{12}$. Similarly, $U_{43}=U_{12}$. Continuing in this vein, and noting that $U$ is not a BCPU, we get that $U$ has Schmidt rank two. It is a contradiction and thus there exists a $V$ of Schmidt rank three.

Up to local permutation matrices, the three nonzero blocks of $V$ are $U_{11}=I_r \op 0_{d_B-r}$,  $U_{12}$,  and $U_{21}$. Up to local permutations on $\cH_A$ we may assume that $U_{23}$ is the other nonzero block in the second big column of $U$. The first $r$ rows of $U_{12}$ are zero, and the first $r$ columns of $U_{21}$ are zero. Since $V$ is of Schmidt rank three, we can find four integers $s,t,u,v$, such that $s>r$ and $v>r$, and $\proj{1}$, $\ketbra{s}{t}$, and $\ketbra{u}{v}$ are pairwise different entries and respectively belong to $U_{11}$,  $U_{12}$,  and $U_{21}$. We choose a nonnormalized input state $\ket{\ps}=(\ket{11}+\ket{22})_{A R_A}\ox[\ket{11}+(1-\d_{1,t})(1-\d_{v,t})\ket{tt}+\ket{vv}]_{BR_B}$. The corresponding output state is $U\ket{\ps}=\frac{1}{2} \ket{11}_{A R_A}\ox\ket{b_1}_{BR_B}+\frac{1}{2} \ket{12}_{A R_A}\ox\ket{b_2}_{BR_B} + \frac{1}{2} \ket{21}_{A R_A}\ox\ket{b_3}_{BR_B} + \frac{1}{2} \ket{32}_{A R_A}\ox\ket{b_4}_{BR_B}$. 

In the cases $t=1$ or $v=t$, we have $\braket{b_1}{b_2}=\braket{b_1}{b_3}=\braket{b_2}{b_3}=0$, and $\ket{b_4}$ may be equal to $\ket{b_1}$ or $\ket{b_3}$, or orthogonal to all of $\ket{b_1}$, $\ket{b_2}$, and $\ket{b_3}$. Thus, the entanglement of the output state is $2$ ebits or $-2\times \frac{1}{4}\log_2 \frac{1}{4}- \frac{1}{2}\log_2 \frac{1}{2}=1.5$ ebits.

The remaining case is that $t,v,1$ are three distinct integers. Since $U$ is a permutation matrix, we have $\braket{b_1}{b_2}=\braket{b_1}{b_3}=\braket{b_2}{b_4}=0$, but $\braket{b_2}{b_3}$ may be $0$ or $\frac{1}{\sqrt{2}}$, since we may always choose the integers $s,t,u,v$ such that $\braket{b_2}{b_3}\ne 1$, and we indeed make such choices here to maximize the output entanglement. An example for the case $\braket{b_2}{b_3}=\frac{1}{\sqrt{2}}$ is given by $r=1$, $U_{12}=0_1 \op 1_1 \op 0_1$, and $U_{21}=0_1\op 1_2$, where ${\rm x}_k$ is ${\rm x}$ times the identity matrix of order $k$; in such case $t=2$ and $v=3$. The case that $\braket{b_2}{b_3}=0$ would give rise to an output entanglement which is too large, thus, in the following we assume $\braket{b_2}{b_3}=\frac{1}{\sqrt{2}}$. Under such condition, it is not hard to show that $\braket{b_1}{b_4}$ and $\braket{b_3}{b_4}$ may be $0$, $\frac{1}{2}$, $\frac{1}{\sqrt{2}}$, or $1$.

Since $U_{11}$ and $U_{21}$ are partial permutation matrices and do not have a common nonzero column, if one of the two quantities $\braket{b_1}{b_4}$ and $\braket{b_3}{b_4}$ is equal to $1$, the other must be $0$. Among the possible cases, the case with the smallest entanglement of the output state is when $\braket{b_2}{b_3}=\frac{1}{\sqrt{2}}$, and one of $\braket{b_1}{b_4}$ and $\braket{b_3}{b_4}$ is $1$ (and the other is $0$). The entanglement of the output state is $H(\frac{1}{4},\frac{3+\sqrt{5}}{8},\frac{3-\sqrt{5}}{8})>1.223$ ebits in this case, where $H(\{x_1,\dots,x_n\}):=-\sum_{j=1}^n x_j \log_2 x_j$ is the entropy function.

In summary, $K_E(U)>1.223$ ebits and the claim holds.
\epf

\section{The proof of Proposition~\ref{pp:aep,control}}
\label{app:{pp:aep,control}}

\bpf
(i) The last inequality in \eqref{eq:K_aep} is obtained by the definition of $K_{Ea}(U)$ and $K_E(U)$. Let us prove the equality in \eqref{eq:K_aep}.
Let $\r_{AR_ABR_B}=\proj{\ps}_{AR_ABR_B}$. We have
\bea
&&
E(U(\ket{\psi}))-E(\ket{\psi})
\notag\\
&=&
S\bigg[\tr_{A R_A}
\bigg(
(\sum^{m}_{j=1} P_j \ox U_j)_{AB}
(\r_{A R_A B R_B})\notag\\
&&(\sum^{m}_{k=1} P_k \ox U_k^\dg)_{AB}
\bigg)
\bigg]
-S(\r_{BR_B})
\notag\\
&=&
S(\sum^{m}_{j=1}
U_j M_j U_j^\dg)
-
S(\r_{BR_B}),
\eea
where $M_j=\tr_{AR_A} \bigg( (P_j)_A \r_{A R_A B R_B} \bigg)$, $\forall j$. Hence $\sum_j M_j = \r_{BR_B}$ and each $M_j$ is a positive semidefinite matrix. Since $\r_{BR_B}$ is arbitrary, we obtain the equality in \eqref{eq:K_aep} by the definition of $K_{Ea}(U)$.

It remains to prove the first inequality in \eqref{eq:K_aep}. We present two different proofs. The first is simpler but the second proof is useful in the proof of (ii) below. The first proof is that the controlled unitaries with $m$ terms can be implemented using a simple protocol in \cite{ygc10} using a maximally entangled state of Schmidt rank $m$, which contains $\log_2 m$ ebits; thus, $E_c(U)\le \log_2 m$, and from \eqref{eq:ecbound} we obtain $K_{Ea}(U)\le \log_2 m$. The second proof is as follows. We use the quantity in the third line of \eqref{eq:K_aep} in place of $K_{Ea}(U)$. Let $M_j'=U_j M_j U_j^\dg$ for $j=1,\cdots,m$. We have
\bea
\label{eq:ssum}
&&
S(\sum_j U_j M_j U_j^\dg)-S(\r)
\notag\\
&\le&
S(\sum_j M_j')-\sum_j \tr M_j \cdot S({M_j \over \tr M_j})
\notag\\
&=&
S(\sum_j \tr M_j' {M_j' \over \tr M_j'} )-\sum_j \tr M_j' \cdot S({M_j' \over \tr M_j'})
\notag\\
&\le&
H(\{\tr M_j'\}).
\eea
The first inequality follows from the concavity of von Neumann entropy and $\r=\sum_j M_j = \sum_j {\tr M_j} \cdot {M_j \over \tr M_j}$. The equality in \eqref{eq:ssum} holds because the von Neumann entropy is invariant under unitary operations. The second inequality in \eqref{eq:ssum} follows from the first inequality in \eqref{eq:concave}, and the observation that $\{\tr M_j'\}$ is a probability distribution. Since $j=1,\cdots,m$, we have $H(\{\tr M_j'\})\le \log_2 m$ and the first inequality in \eqref{eq:K_aep} holds.

(ii) Suppose the first inequality in \eqref{eq:K_aep} becomes the equality. It is equivalent to the condition that both inequalities in \eqref{eq:ssum} become equalities, and $H(\{\tr M_j'\}) = \log_2 m$. It implies that any $M_j'$ is nonzero.
Lemma \ref{le:concave} implies that $M_j\propto\r$ for any $j$, and $M_j'M_k'=0$ for $j\ne k$. Since $H(\{\tr M_j'\}) = \log_2 m$ and $M_j'=U_j M_j U_j^\dg$, we have $M_j={1\over m}\r$. Thus,
\bea
\label{eq:rujuk}
\r (U_j^\dg U_k \ox I_{R_B}) \r=0
\eea
for any $j,k$ and $j\ne k$. Since $\r$ is a mixed state, we can project it onto a pure state in the Schmidt decomposition, namely $\ket{\ps}=\sum_i \sqrt{c_i}\ket{a_i,b_i}$. Then \eqref{eq:rujuk} becomes $\sum_i c_i \bra{a_i}U_j^\dg U_k\ket{a_i}=0$. Setting $\s=\sum_i c_i \proj{a_i}$
implies the ``only if'' part except that $j<k$ is also allowed. It can be excluded because $\tr (\s U_j^\dg U_k )=0$ is equivalent to $\tr (\s U_k^\dg U_j )=0$.
On the other hand the ``if'' part follows by assuming $M_j={1\over m}\r$ for $j=1,\cdots,m$.

To prove the last-but-one assertion, if $U_i$ are all diagonal then so are $U_j^\dg U_k$. Since $\tr (\s U_j^\dg U_k)=0$ for all $j\ne k$, we have
$\tr (\s' U_j^\dg U_k)=0$ for all $j\ne k$, where $\s'$ is the diagonal matrix whose diagonal entries are the same as those of $\s$. So $\s'$ is still a quantum state.

To prove the last assertion, if $U_i$ are all real, then $\tr (\s^* U_j^\dg U_k)=0$ for all $j\ne k$. The sum of this equation and $\tr (\s U_j^\dg U_k)=0$ implies $\tr (\s' U_j^\dg U_k)=0$ where $\s'={\s+\s^*\over2}$ is real.

(iii) The last claim follows from the first claim. If $K_E(U)=\log_2 m$, then $K_{Ea}(U)=\log_2 m$ by (i). It suffices to prove the statement that $K_{Ea}(U)=\log_2 m$ implies $K_E(U)=\log_2 m$. The statement follows from (ii) and Lemma \ref{le:control} (i).

(iv) The statement follows from assertion (i) and Lemma \ref{le:control} (i).
This completes the proof. \epf

\bibliography{channelcontrol}

\end{document}